\documentclass[12pt]{iopart}

\usepackage{amsfonts,bm}
\usepackage{graphicx}
\usepackage{longtable}
\usepackage{epsfig}
\usepackage{color}
\usepackage{arydshln}
\usepackage{cite}

\def\beq{\begin{equation}}
\def\eeq{\end{equation}}
\def\beqa{\begin{eqnarray}}
\def\eeqa{\end{eqnarray}}

\newcommand{\myfrac}[2]{\leavevmode\kern.1em
   \raise.5ex\hbox{\the\scriptfont0 #1}\kern-.1em
   /\kern-.15em\lower.25ex\hbox{\the\scriptfont0 #2}}
 \renewcommand{\tr}{\hbox{Tr}}

\begin{document}

\title[Phase-space and mutually orthogonal Latin squares]{Discrete
phase-space approach to mutually orthogonal Latin squares}
\author{Mario Gaeta$\,^{1}$, Olivia Di Matteo$\,^{2}$\footnote{Current address:
Department of Physics and Astronomy, Institute for Quantum Computing, University of Waterloo} , 
Andrei B. Klimov$\,^{1}$ and Hubert de Guise$\,^{2}$}

\begin{abstract}
We show there is a natural connection between Latin squares and commutative
sets of monomials defining geometric structures in finite
phase-space of prime power dimensions. A complete  set of such monomials
defines a mutually unbiased basis (MUB) and may be associated with a complete
set of mutually orthogonal Latin squares (MOLS). We translate some possible
operations on the monomial sets into isomorphisms of Latin squares, and find
a general form of permutations that map between Latin squares corresponding
to unitarily equivalent mutually unbiased sets.
\end{abstract}


\address{$\,^1$ Departamento de F\'{\i}sica, Universidad de Guadalajara, 44420 Guadalajara, Jalisco, Mexico } 
\address{$\,^2$ Department of Physics, Lakehead University, Thunder Bay, Ontario P7B 5E1, Canada}

%\maketitle

\section{Introduction}

Relations between mutually orthogonal Latin squares (MOLS) \cite{laywine_mullen,denes_keedwell1}  and
mutually unbiased bases (MUBs) \cite{wocjan_beth, klappenecker_rottler,
wootters_solo, zauner} have been the subject of renewed interest \cite{paterek1, hallrao1, hallrao2, paterek2,Ghiu}. 
MOLS have been studied since Euler; they find applications in the
design of experiments \cite{Fischer}, coding theory (see for instance the
text on Latin squares by \cite{denes_keedwell1}), compressed sensing \cite{Herman,Flammia} and a
variety of areas in pure and applied mathematics \cite{Colbourn}. 
Mutually unbiased bases \cite{Wootters,review paper} on the other hand,
have a much shorter history; a complete set of MUBs constitutes an
optimal experimental choice for reconstructing the density matrix of a
system, a property that strongly suggests a connection between MUBs and MOLS.

One (of many possible) way of obtaining a complete set of MUBs is based on
the construction of eigenstates of disjoint sets of commuting monomials 
\cite{klappenecker_rottler, mub, lawrence,multicomp}. In prime power dimensions, the explicit construction of such sets 
can be carried out in terms of symplectic spreads \cite{Calderblanks, kantor} or planar functions \cite{scott}. 
There is a simple correspondence between these types of MUBs and MOLS \cite{wocjan_beth,
paterek1, hallrao1, hallrao2, paterek2}. One concludes from this
correspondence that sets of commuting monomials can be nicely represented as
particular geometrical structures in a finite phase-space \cite{curvepapers}. This opens up
a possibility of connecting phase-space geometry with MOLS.

Here we focus on the relation between MOLS and MUBs from the perspective of
phase-space. In particular, we analyze what types of MUBs can be directly
converted into MOLS, and find the image of some useful
Clifford transformations of the MUBs on the corresponding MOLS. We also discuss the factorization structure of MUBs 
\cite{lawrence, romero} on the level of MOLS: it will be
shown that for MUBs associated to MOLS, a set of ``legal'' transformations
on MUBs (comprising transformations of the CNOT type on
multi-particle MUBs, plus some specific local transformations) induce isomorphisms between the corresponding MOLS. In
particular, starting with MUBs associated with Desarguesian MOLS, ``legal''
transformations will produce a new set of MUBs also associated to
Desarguesian MOLS. This leads to the observation that, although CNOT
operations change the separability properties of MUBs, they do not affect
the type of MOLS.

Finally, we analyze the inverse relation between MOLS and MUBs of monomial
type and propose an explicit procedure to identify MOLS that are related to such MUBs.

\section{MUBs, monomials, and commutative curves}\label{section2}

We start by enumerating elements in the field $\mathbb{F}_{p^{n}}$ as 
\begin{equation}
\mathbb{F}_{p^{n}}=\{\sigma ^{i},i=0,\ldots ,p^{n}-1\}\,,
\label{fieldelements}
\end{equation}
with
\begin{equation}
 \sigma^{i} = \left\{
 \begin{array}{ll}
0 & \hbox{ if } i = 0\, , \\ 
\sigma^i & \hbox{ if } i = 1, \ldots, p^n - 1 \, ,
\end{array}
\right. \label{sigmaiconvention}
\end{equation}
where $\sigma \,\in \,\mathbb{F}_{p^{n}}$ is a primitive element (a root of a minimal irreducible polynomial).  
\emph{We alert the reader to our unusual notation where $\sigma^0=0$:  
this choice is very convenient as we will construct Latin squares from the exponents of a primitive element in the finite field
$\mathbb{F}_{p^{n}}$.}
Generic elements in $\mathbb{F}_{p^{n}}$ are denoted by $\alpha $ and 
$\beta $. 
It is sometimes convenient to think of $\alpha $ and $\beta $ as parametric functions on $\mathbb{F}_{p^{n}}$, in which case 
we write $\alpha (\sigma ^{i})$ and $\beta (\sigma^{i})$. 
All arithmetic is done over the finite field.

\subsection{Monomials}

To each $\sigma^i \in \mathbb{F}_{p^{n}}$ we associate a ket vector 
$\vert \sigma^i \rangle$ so that $\{|\sigma^i \rangle ,\, i=0,\ldots,p^n-1\}$ is an orthonormal basis in the Hilbert space of 
an $n$ qudit system: $\langle \sigma^i |\sigma^j \rangle =\delta _{ij}$.

We introduce two families of basic operators $\{Z_{\alpha} , \alpha\in 
\mathbb{F}_{p^{n}}\}$ and $\{X_{\beta}, \beta\in \mathbb{F}_{p^{n}}\}$,
conveniently taken to be of the form 
\begin{equation}
Z_{\alpha}=\sum_{i=0}^{p^n-1}\chi (\alpha\sigma^i )\,|\sigma^i \rangle
\langle \sigma^i |\, ,\qquad X_{\beta }=\sum_{i=0}^{p^n-1}|\sigma^i
+\beta\rangle\langle \sigma^i |\,,
\end{equation}
where 
\begin{equation}
\chi (\alpha )=\exp [\frac{2\pi i}{p}\tr(\alpha )]\, , \quad \tr(\alpha
)=\alpha +\alpha ^{p}+...+\alpha ^{p^{n-1}}\ \ (\hbox{mod}\ p)\, ,
\end{equation}
and $\tr$ is the usual trace mapping $\mathbb{F}_{p^{n}}\to \mathbb{F}_{p}$ 
\cite{multicomp}. Note this implies $Z_0=X_0\equiv\mathinner{\hbox{1}%
\mkern-4mu\hbox{l}}$.

It will prove extremely useful to consider $\mathbb{F}_{p^{n}}$ as an $n$-dimensional linear space, so that any $\alpha \in \mathbb{F}_{p^{n}}$ can be
expressed as a linear combination of elements of the (almost) self-dual
basis $\left\{ \theta _{1},...,\theta _{n}\right\} $: 
\begin{eqnarray}
&&\alpha =\sum_{i=1}^{n}a_{i}\theta _{i},\quad a_{i}\in \mathbb{Z}_{p},
\qquad \beta =\sum_{i=1}^{n}b_{i}\theta _{i},\quad b_{i}\in \mathbb{Z}_{p}\,
,  \nonumber \\
&&\tr(\theta _{i}\theta_{j})=c_{j}\delta _{ij},\ c_{j}\in \mathbb{Z}_{p}\, .
\end{eqnarray}
It is always possible to enumerate the basis elements $\{\theta_i\}$ in an
order where $c_j=1$ for $j>1$. We can then write 
\begin{equation}
Z_{\alpha }=\mathcal{Z}^{c_1a_{1}}\otimes \ldots \otimes \mathcal{Z}^{c_n
a_{n}},\quad X_{\beta}={\mathcal{X}}^{b_{1}}\otimes \ldots \otimes \mathcal{X}^{b_{n}},
\end{equation}%
where $\mathcal{Z}$ and $\mathcal{X}$ are the generalized $p-$dimensional
Pauli matrices. The monomials 
\begin{equation}
Z_{\alpha}X_{\beta}={\mathcal{Z}}^{c_1a_1}{\mathcal{X}}^{b_1}\otimes {\mathcal{Z}}^{c_2 a_2}{\mathcal{X}}^{b_2}\otimes\ldots 
\otimes {\mathcal{Z}}^{c_n a_n}{\mathcal{X}}^{b_n}  \label{basicmonomials}
\end{equation}
are then elements of the generalized Pauli group $\mathcal{P}_{n}$, \textit{i.e.} satisfy 
$Z_{\alpha }X_{\beta }=\chi(\alpha \beta )\,X_{\beta}Z_{\alpha }$.

\subsection{ Additive and commutative curves}

First, recall that two orthonormal bases $\{|A_{i}\rangle ,i=1,\ldots ,d\}$
and $\{|B_{j}\rangle ,j=1,\ldots ,d\}$ are said to be mutually unbiased if 
\begin{equation}
|\langle \,A_{i}\,|\,B_{j}\,\rangle |=\frac{1}{\sqrt{d}}\,,\quad \forall
i,j\,.  \label{mubcondition}
\end{equation}
The set $\{|A_{i}\rangle \}$ can taken as  the common eigenvectors of $d-1$
commuting operators $\{|A_{i}\rangle \langle A_{i}|\}$, while 
$\{|B_{j}\rangle \}$ is constructed from another set of commuting operators,
disjoint from the set used to construct $\{|A_{i}\rangle \}$. Two sets of
commuting operators are traditionally referred to as being mutually unbiased
if they have eigenvectors satisfying the condition of Eq. (\ref{mubcondition}).

If $d+1$ disjoint mutually unbiased sets of commuting operators exist, we
have a complete set of MUBs. With $d=p^{n}$ and $p$ a prime, it is known
that (in general, several) complete sets of MUBs exist. Here we will focus
on complete sets of commuting \textit{monomials }\cite{mub, multicomp} of the form given in Eq.(\ref{basicmonomials}). Such
sets can be:

\begin{itemize}
\item[a)] unitarily equivalent, but not locally equivalent; these sets are
distinguished by their factorization structure \cite{lawrence, romero},

\item[b)] unitarily inequivalent \cite{kantor}.
\end{itemize}

It is convenient to label sets of commuting monomials by points of additive,
commutative curves  \cite{curvepapers} $\{Z_{\alpha (\sigma^{i})}X_{\beta (\sigma ^{i})},i=1,...,p^{n}-1\}$ in 
discrete phase-space. For a system of $n$ qudits phase-space is a discrete grid of $p^{n}\times p^{n}$ 
points~$\{(\alpha ,\beta ),\alpha ,\beta \in \mathbb{F}_{p^{n}}\}$ 
\cite{Wootters, Gibbons, Vourdas:2007dq}, whose axes are labelled by
elements of the finite field $\mathbb{F}_{p^{n}}$, endowing the grid with
standard geometrical properties~\cite{Lidl:1986fk}. With this geometrical
structure, an operator $Z_{\alpha }X_{\beta }$ is mapped to a unique point
in phase-space.

If the points $(\alpha ,\beta )$ of a curve are given in parametric form 
\begin{equation}
\alpha =\alpha (\sigma ^{i})\,,\qquad \beta =\beta (\sigma ^{i})\,,
\end{equation}%
then \emph{additive curves} satisfy the requirement: 
\begin{equation}
\alpha (\sigma ^{i}+\sigma ^{j})=\alpha (\sigma ^{i})+\alpha (\sigma
^{j}),\qquad \beta (\sigma ^{i}+\sigma ^{j})=\beta (\sigma ^{i})+\beta
(\sigma ^{j}),  \label{additivity}
\end{equation}%
for any $\sigma ^{i},\sigma ^{j}\in \mathbb{F}_{p^{n}}$. To enforce the
commutativity of operators within  a set, namely 
\begin{equation}
\lbrack Z_{\alpha (\sigma ^{i})}X_{\beta (\sigma ^{i})},Z_{\alpha (\sigma
^{j})}X_{\beta (\sigma ^{j})}]=0,
\end{equation}
we must consider \emph{commutative curves}, by which we
understand such curves satisfy 
\begin{equation}
\tr(\alpha (\sigma ^{i})\beta (\sigma ^{j}))=\tr(\alpha (\sigma ^{j})\beta
(\sigma ^{i}))\,.  \label{commutativity}
\end{equation}

\section{Latin squares and commutative curves}

\label{lsandcommutativecurves}

\subsection{Invertibility and unbiasedness}

We focus on commutative curves defined by invertible functions.
Invertibility means there is a one-to-one correspondence between coordinates 
$\alpha$ and $\beta $ on the curve; alternatively, no particular value $%
\alpha(\sigma^i)$ or $\beta (\sigma^i )$ occurs more than once in a given
curve, so $\alpha(\sigma^i)$ and $\beta (\sigma^i )$ are just the field
elements enumerated in some order generally different from that given in Eq.~(\ref{fieldelements}).

Hence, points on an invertible curve can be written in the form $\beta=f(\alpha )$ where $f(\alpha )$ is a non-singular 
(invertible) function such that
\begin{equation}
f(\alpha )=\sum_{i=0}^{n-1}\phi _{i}\,\alpha ^{p^{i}},\quad 
\phi _{k}:=\phi_{n-k}^{p^{k}}\, , k=1,\ldots ,[(n-1)/2]\,, 
\end{equation}
and $\phi _{i},\alpha \in \mathbb{F}_{p^{n}}$.
Here,  $[\,]$ denotes the integer part.   If $n$ is even, there is the additional requirement
$\phi _{n/2}=\phi _{n/2}^{p^{n/2}}$.
\cite{curvepapers,graphstates} We can thus also write $\alpha =f^{-1}(\beta )$. 

Let us recall that a  (general) Latin square is a $d\times d$ array where the symbols 
$0,\ldots,d- 1$ occur once and only once in each row
and each column. Two Latin squares $L^{(1)}$ and $L^{(2)}$ are mutually
orthogonal if all the pairs $(L_{ij}^{(1)},L_{ij}^{(2)}),i,j=0,\ldots d-1$, occur once and only once.

It is straightforward to see that to each \textit{invertible} curve $\beta=f(\alpha)$ corresponds a Latin square with entries 
\begin{equation}
L_{ij}^{(f)}=\sigma ^{j}+f(\sigma ^{i})\equiv \sigma ^{k},\quad
i,j=0,...,N-1,  \label{basicLS}
\end{equation}
where, as indicated in Eq.~(\ref{sigmaiconvention}), $\sigma ^{k}$ is the $k$'th power of a primitive element 
$\sigma \,\in \,\mathbb{F}_{p^{n}}$. For specified $f$ and all $i,j$, Eq. (\ref{basicLS}) produces some other element 
$\sigma ^{k}$ in the field; for simplicity, the entry $L_{ij}^{(f)}$ at position $(i,j)$ of the Latin square will be written as $k$. 
The LS constructed according to (\ref{basicLS}) is \emph{standard}, since the symbols of the first row are ordered in increasing 
powers of $\sigma $.

As a simple illustration of Eq.~(\ref{basicLS}) we consider a two-qubit system, for which the relevant field is $\mathbb{F}_{2^{2}}$, 
with elements constructed using the irreducible polynomial $\sigma ^{2}+\sigma +1$.  Choose the function 
$\beta =f(\sigma ^{i})=\sigma \sigma ^{i}$ (for instance).  Indexing rows and columns from 0, the resultant square is 
\begin{equation}
L^{(\sigma \alpha )}=\left( {\renewcommand{\arraystretch}{0.85}%
\renewcommand{\arraycolsep}{2.95pt}{\small 
\begin{array}{cccc}
0 & 1 & 2 & 3 \\ 
2 & 3 & 0 & 1 \\ 
3 & 2 & 1 & 0 \\ 
1 & 0 & 3 & 2
\end{array}}}\right) \,.
\end{equation}

Disjoint sets of commuting monomials $\{Z_{\alpha }X_{f_{\xi }(\alpha)},\xi=0,1,\ldots ,p^{n}-2\}$ are mapped to curves with no point 
in common (except at the origin) \cite{curvepapers}.

For example, one easily verifies that the set of operators $%
\{Z_{\sigma^{i}}X_{\lambda \sigma ^{i}},i=0,\ldots p^{n}-1\}$, for some
fixed $\lambda\in \mathbb{F}^*_{p^n} \equiv \mathbb{F}_{p^{n}} \backslash
\{0 \} $, commute with each other. 
The set of points $\{(\sigma ^{i},\lambda \sigma ^{i})\}$ is a straight line
(a ray) with slope $\lambda $ in discrete phase-space.

The connection between MUBs and MOLS is that, under the proper conditions,
the same ``bundle'' of nonintersecting curves $\{f_{\xi }\}$ is used to
simultaneously construct a complete set of MUBs and MOLS: to a set of MUBs
described by $p^{n}-1$ invertible, non-intersecting curves corresponds a
complete set of MOLS. Since the curves in the bundle do not intersect
(except at the origin), the associated Latin squares will be orthogonal. The
maximum number of invertible curves in a bundle describing a complete set of
MUBs is $p^{n}-1$. 
%but not all the possible sets of MUBs necessarily contain this many. 
(It would appear there are two curves missing, as we need $p^{n}+1$ sets of
commuting operators, but only have $p^{n}-1$ invertible curves. This is
because two of the curves are always non-invertible. For instance, the curve 
$\beta =0$ corresponding to operators of the type $Z_{\alpha}$, and the curve $\alpha =0$ corresponds to operators of 
the type ${X}_\beta$, are not invertible. The squares corresponding
to these curves have identical entries across each column and row,
respectively. These special curves and resulting squares are excluded from
our discussion.)

\subsection{The adjacency matrix}

We can use the (almost) self-dual basis $\left\{ \theta_{1},...,\theta_{n}\right\} $ to write Eq.~(\ref{basicLS}) in a compact form
useful for later analysis.
Let us expand 
\begin{equation}
\sigma ^{i}=\sum_{k}s_{k}^{i}c_{k}^{-1}\theta _{k}\,,\qquad \qquad s_{k}^{i}=%
\tr[\sigma^i\theta_k]\,,  \label{sigmaiexpand}
\end{equation}%
with $c_i$ given in Eq. (5), and define
\begin{equation}
\mathbf{s}^{i}=\left( s_{1}^{i},\ldots, s_{n}^{i} \right) \,,\quad %
\bm{\theta}=\left( 
\begin{array}{c}
\theta _{1} \\ 
\vdots \\ 
\theta _{n}%
\end{array}%
\right) \,,\quad \mathbf{C}=\left( 
\begin{array}{ccc}
c_{1} & \ldots & 0 \\ 
0 & \ddots & \vdots \\ 
0 & \ldots & c_{n}%
\end{array}%
\right) \,.  \label{adjmatrixdefinitions}
\end{equation}

As mentioned at the start of Sec. \ref{section2}, the Latin squares are not constructed
from the element $\sigma^i$ but rather from its exponent. Thus, in
transforming Latin squares we are ultimately interested in the exponent $i$
of the vector $\mathbf{s}^i$ associated with the element $\sigma^i$.

We introduce the adjacency matrix \cite{graph states} associated to a curve $f$: 
\begin{equation}
\mathbf{\Gamma }_{k\ell }^{(f)}=\tr\left( c_{\ell }^{-1}\theta _{\ell
}\,f(c_{k}^{-1}\theta _{k})\right) \,\in \mathbb{F}_{p}\,.
\end{equation}

The adjacency matrix has a number of useful properties:

\begin{itemize}
\item[a)]For invertible curves, det$[\mathbf{\Gamma}^{(f)}]\neq 0$

\item[b)]
A necessary and sufficient condition for an adjacency matrix to describe an additive, commutative curve $f$ is that
it be symmetric: $\mathbf{\Gamma}^{(f)}_{k\ell}=\mathbf{\Gamma}^{(f)}_{\ell k}$ \cite{graphstates}. 

\item[c)] For composition of two functions $f$ and $g$, 
\begin{equation}
\mathbf{\Gamma}^{(f\circ g)} = \mathbf{\Gamma}^{(g)} \,\mathbf{C} \,\mathbf{%
\Gamma }^{(f)}  \label{adjmatcomposition}
\end{equation}

\item[d)] For the inverse of a curve, $f^{-1}$, 
\begin{equation}
\mathbf{\Gamma }^{(f^{-1})}=\mathbf{C}^{-1} (\mathbf{\Gamma }^{(f)})^{-1}\,%
\mathbf{C}^{-1}\,  \label{adjmatinversion}
\end{equation}

\end{itemize}

In addition, if $f$ is the identity function $\mathrm{Id}$ such that $%
\mathrm{{Id}(\sigma ^{i})=\sigma ^{i}}$, then $\mathbf{\Gamma }^{(\mathrm{{Id})}}=\mathbf{C}^{-1}$. 
It follows from Eqs.~(\ref{sigmaiexpand}) - (\ref{adjmatrixdefinitions}) that 
\begin{equation}
\sigma ^{j}=\mathbf{s}^{j}\,\mathbf{C}^{-1}\bm{\theta}
=\mathbf{s}^{j}\, \mathbf{\Gamma}^{(\mathrm{{Id})}}\,\bm{\theta}\,,\qquad f(\sigma ^{i})
=\mathbf{s}^{i}\,\mathbf{\Gamma }^{(f)}\,\bm{\theta}\,,
\end{equation}
and we can rewrite Eq.(\ref{basicLS}) quite compactly as the matrix product 
\begin{equation}
L_{ij}^{(f)}=\left( \mathbf{s}^{j}\, \mathbf{C}^{-1}+\mathbf{s}^{i}\,\mathbf{\Gamma }^{(f)}\right) \bm{\theta }.  \label{OLSG}
\end{equation}

\section{Standard and non-standard Latin squares}

The points of a curve $f$ need not necessarily be listed in increasing
powers of $\sigma $ but can also be given in parametric form $(\alpha (\sigma ^{i}),\beta (\sigma ^{i}))$ where both 
$\alpha (\sigma ^{i})$ and 
$\beta (\sigma ^{i})$ are invertible functions, so that $f(\sigma ^{i})=\beta (\alpha ^{-1}(\sigma ^{i})).$  We make the 
important  observation that the
parametric $(\alpha (\sigma ^{i}),\beta (\sigma ^{i}))$ and explicit $f(\sigma ^{i})$ forms of a curve differ only
by the order in which points are enumerated.  We need to use both forms as we will show 
that different orderings correspond to Latin squares differing by a permutation of columns.

If we consider the adjacency matrices $\Gamma ^{(\alpha )}$ and $\Gamma
^{(\beta )}$ of a parametric curve $(\alpha (\sigma ^{i}),\beta (\sigma
^{i}))$, we obtain from Eq.~(\ref{adjmatcomposition}) 
\begin{equation}
\mathbf{\Gamma }^{(f)}=\left( \mathbf{\Gamma }^{(\alpha )}\mathbf{C}\right)
^{-1}\mathbf{\Gamma }^{(\beta )}  \label{Gf_as_GaGb}
\end{equation}

A LS can be constructed using this ordering: 
\begin{equation}
\tilde{L}^{(f)}_{ij}=\left( \mathbf{s}^j\,\mathbf{\Gamma }^{(\alpha)} +\mathbf{s}^i\,\mathbf{\Gamma }^{(\beta)}\right) 
\bm{\theta,}  \label{DLSG}
\end{equation}
but it is not standard. We use $\tilde{L}$ to denote such a non-standard square. 

The columns and rows of $\tilde{L}^{(f)}_{ij}$ are related to those of
standard LS of Eq.~(\ref{OLSG}) as follows. Consider the very first row, $i = 0$, and compare the ordering of the symbols in 
this row for both $\tilde{L}$ and $L$: the entry $(0,j)$ of $\tilde{L}^{(f)}$ contains symbol $k$ such that 
\begin{equation}
\tilde{L}^{(f)}_{0j}=\alpha(\sigma^j)=\mathbf{s}^j\,\bm{\Gamma}^{(\alpha)}%
\bm{\theta} =\mathbf{s}^k\mathbf{C}^{-1}\bm{\theta}\qquad\hbox{for some $k$.}
\end{equation}
Thus, 
%sending column $j$ of $\tilde{L}^{(f)}$ to column $k$, with $k$ obtained from
the column permutation 
\begin{equation}
\hbox{column $j\to k$ with $k$ such that } \quad \mathbf{s}^k=\mathbf{s}^j\,
\bm{\Gamma}^{(\alpha)}\mathbf{C}  \label{columnpermutation}
\end{equation}
will bring $\tilde{L}^{(f)}$ to a standard square which still differs from $L^{(f)}$ by row permutations. 
Now, set $j=0$ in $\tilde L^{(f)}$ so the entries of the first column are 
\begin{eqnarray}
\tilde L^{(f)}_{i0}=\beta(\sigma^i)=\mathbf{s}^i\,\bm{\Gamma}^{(\beta)}
\bm{\theta} &=&\mathbf{s}^k\bm{\Gamma}^{(f)}\bm{\theta}  \nonumber \\
&=&\mathbf{s}^k\left(\bm{\Gamma}^{(\alpha)}\mathbf{C}\right)^{-1} \bm{\Gamma}
^{(\beta)}\bm{\theta}\quad\hbox{for some $k$,}
\end{eqnarray}
where Eq.~(\ref{Gf_as_GaGb}) has been used. Thus, the row permutation 
\begin{equation}
\hbox{row $i\to k$ with $k$ such that} \quad \mathbf{s}^k=\mathbf{s}^i\left(%
\bm{\Gamma}^{(\alpha)}\mathbf{C}\right) ,  \label{rowpermutation}
\end{equation}
combined with the column permutation of Eq.~(\ref{columnpermutation}) will
bring $\tilde L^{(f)}$ to $L^{(f)}$. We note here that the permutations of
row and columns commute.

\section{Operations on individual monomials and corresponding Latin squares}

Since any unitary transformation applied to an $n$-qudit state can be
decomposed into a product of CNOT-type operations and local transformations \cite{gates},
we consider the result of these operations on monomials and the
corresponding permutations of rows, columns and symbols of Latin squares
such operations induce. For this purpose, it is simplest to initially list
fields elements in increasing powers of $\sigma$. 
%as this results, following Eq.~(\ref{OLSG}), in a standard Latin square.

\subsection{CNOT transformations}

The standard CNOT operation performed on qudits $p$ and $q$ is defined as 
\begin{equation}
\hbox{CNOT}_{pq}\left\vert \lambda \rangle \right. =\left\vert \lambda 
+\Tr\left( \lambda c_{p}^{-1}\theta _{p}\right) \theta _{q}\rangle \right. ,\qquad \lambda \in \mathbb{F}_{p^{n}}\,,
\end{equation}%
so that under the action of the $m$-th power CNOT 
(with $m=0,1,\ldots ,p-1$)
\begin{equation}
\hbox{CNOT}_{pq}^{m}\left\vert \lambda \rangle \right. =\left\vert \lambda
+m\,\Tr\left( \lambda c_{p}^{-1}\theta _{p}\right) \theta _{q}\rangle
\right. , 
\end{equation}
the monomials are transformed as 
\begin{eqnarray}
&&\ldots \otimes Z_{p}^{c_{p}a_{p}}X_{p}^{b_{p}}\otimes \ldots \otimes
Z_{q}^{c_{q}a_{q}}X_{q}^{b_{q}}\otimes \ldots  \nonumber \\
&&\qquad \rightarrow \ldots \otimes
Z_{p}^{c_{p}a_{p}-m\,c_{q}a_{q}}X_{p}^{b_{p}}\otimes \ldots \otimes
Z_{q}^{c_{q}a_{q}}X_{q}^{m\,b_{p}+b_{q}}\otimes \ldots \,.
\end{eqnarray}%
In terms of field elements, $Z_{\sigma ^{i}}X_{f(\sigma ^{i})}$ is
transformed into $Z_{\alpha (\sigma ^{i})}X_{\beta (\sigma ^{i})}$, where 
\cite{graphstates} 
\begin{eqnarray}
\alpha (\sigma ^{i}) &=&\sigma ^{i}-m\,\hbox{Tr}\left( \sigma ^{i}\theta
_{q}\right) c_{p}^{-1}\theta _{p}\,,  \nonumber \\
\beta (\sigma ^{i}) &=&f(\sigma ^{i})+m\,\hbox{Tr}\left( f(\sigma
^{i})c_{p}^{-1}\theta _{p}\right) \theta _{q}\,,
\end{eqnarray}

The action of CNOT transforms the adjacency matrix of a standard LS to a new adjacency
matrix given by 
\begin{equation}
\mathbf{\Gamma }^{(g)}=\left(\mathbf{X}{{\,}^{m}_{p,q} }\right)^{T}\mathbf{%
\Gamma }^{(f)}\mathbf{X}{{\,}^{m}_{p,q} }\, , \label{CNOT_adjmat}
\end{equation}
where $T$ denotes transposition.
Since det$ [\mathbf{X}^{m}_{p,q}] \ne 0$, the CNOT operation does not change the
invertibility property of a curve.

The Latin square associated with the set $Z_{\alpha (\sigma ^{i})}X_{\beta(\sigma ^{i})}$ is obtained using 
Eq.~(\ref{DLSG}), with new adjacency matrices 
\begin{equation}
\mathbf{\Gamma }^{(\alpha )}=\left( \mathbf{X}{\,}_{p,q}^{-m}\right) ^{T}%
\mathbf{C}^{-1},\quad \mathbf{\Gamma }^{(\beta )}=\mathbf{\Gamma }^{(f)}%
\mathbf{X}{\,}_{p,q}^{m}\, ,  \label{transformedadjmat}
\end{equation}
where we use
\begin{equation}
\left( \mathbf{X}_{p,q}^{m}\right) _{ij}=\delta _{ij}+m\delta _{ip}\delta_{jq}\, ,  \label{CNOTm}
\end{equation}
to represent the CNOT operation. The resulting non-standard LS is 
\begin{eqnarray}
\tilde{L}^{(g)}_{ij} &=& \left( \mathbf{s}^j \left(\mathbf{X}{\,}^{-m}_{p,q}
\right)^{T} \mathbf{C}^{-1} +\mathbf{s}^i\left(\mathbf{X}{\,}^{-m}_{p,q}
\right)^{T} \mathbf{\Gamma}^{(g)}\right) \bm{\theta } , \label{LStildeg}
\end{eqnarray}
where Eq.~(\ref{Gf_as_GaGb}) (with $f\to g$) and $\left(\mathbf{X}%
^{m}_{p,q}\right)^{-1}=\left(\mathbf{X}^{-m}_{p,q}\right)$ have been used.

The Latin square $\tilde L^{(g)}$ of Eq.~(\ref{LStildeg}) is then clearly
related to $L^{(g)}$ by the permutations: 
\begin{equation}
\begin{array}{rl}
\hbox{row $i\to k$ with $k$ such that} & \mathbf{s}^{k} = \mathbf{s}^i
\left(\mathbf{X}{\,}^{-m}_{p,q} \right)^{T} \, , \\ 
\hbox{column $j\to k$ with $k$ such that} & \mathbf{s}^{k} =\mathbf{s}^j\left(\mathbf{X}^{-m}_{p,q} \right)^{T}\, ,
\end{array}
\label{colperm_CNOT_nstos}
\end{equation}

We can also obtain the sequence of permutations taking $\tilde L^{(g)}$ of Eq.~(\ref{LStildeg}) for the curve $g$ to the original, 
standard form LS of $f$, $L^{(f)}$. We can manipulate Eq.~(\ref{LStildeg}) to the form: 
\begin{eqnarray}
\tilde{L}^{(g)}_{ij} &=& \left( \mathbf{s}^j\, \mathbf{W}\,\mathbf{C}^{-1} + 
\mathbf{s}^i\, \mathbf{\Gamma}^{(f)} \right) \left[ \mathbf{X}^{m}_{p,q} \bm{\theta } \right]\, ,
\end{eqnarray}
where 
$\mathbf{W} := \left(\mathbf{X}{\,}^{-m}_{p,q} \right)^{T} \mathbf{C}^{-1} \mathbf{X}{\,}^{-m}_{p,q} \mathbf{C}$. $\tilde{L}^{(g)}$ is thus
transformed back to the original square $L^{(f)}$ by the following permutations: 
\begin{equation}
\begin{array}{rll}
\hbox{no permutation of rows,} & \, &  \\ 
\hbox{column $j\to k$ with $k$ such that } &  & \mathbf{s}^k = \mathbf{s}^j\,%
\mathbf{W}\, , \\ 
\hbox{symbol $k\to j$ with $k$ such that } &  & \mathbf{s}^k = \mathbf{s}^j\,%
\mathbf{C}^{-1}\mathbf{X}^m_{p,q}\mathbf{C} := \mathbf{s}^j\,\mathbf{V}\, .%
\end{array}
\label{colperm_CNOT_to_original}
\end{equation}
The symbol swap is inferred from Eq.~(\ref{sigmaiexpand}): 
in going to the new basis $ \left[ \mathbf{X}^{m}_{p,q} \bm{\theta } \right]$,
\beq
\sigma^j=\mathbf{s}^{j}\,\mathbf{C}^{-1}\,\bm{\theta}\to \mathbf{s}^{j}\,
\mathbf{C}^{-1}\,\mathbf{X}^{m}_{p,q}\,\bm{\theta} = 
\mathbf{s}^{j}\,
\mathbf{V}\,\mathbf{C}^{-1}\,\bm{\theta}=\mathbf{s}^k\,\mathbf{C}^{-1}\,\bm{\theta}\, ,
\eeq
yielding the last of Eq.~(\ref{colperm_CNOT_to_original}).

Write $L^{(g)}$ in the form of Eq. (\ref{OLSG}) and compare it to $L^{(f)}$: 
\begin{eqnarray}
L^{(g)}_{ij} &=& \left( \mathbf{s}^j\, \mathbf{C}^{-1} + \mathbf{s}^i\, 
\mathbf{\Gamma}^{(g)} \right) \bm{\theta}  \nonumber \\
& = & \left( \mathbf{s}^j\, \mathbf{C}^{-1} \mathbf{X}{^{-m}_{p,q} } + 
\mathbf{s}^i \left(\mathbf{X}{^{m}_{p,q} }\right)^{T} \mathbf{\Gamma }%
^{(f)}\right) \mathbf{X}{^{m}_{p,q} } \bm{\theta} \\
& = & \left( \mathbf{s}^j\, \mathbf{V}^{-1}\, \mathbf{C}^{-1} + \mathbf{s}^i \left(%
\mathbf{X}{{\,}^{m}_{p,q} }\right)^{T}\mathbf{\Gamma }^{(f)} \right) \mathbf{%
X}{{\,}^{m}_{p,q} } \bm{\theta}\, .  \nonumber
\end{eqnarray}
Comparing this to the form of Eq. (\ref{OLSG}), we can see that the following transformation brings us from $L^{(g)}$ back to $L^{(f)}$: 
\begin{equation}
\begin{array}{rll}
\hbox{row $i\to k$ with $k$ such that} & \quad & \mathbf{s}^k= \mathbf{s}^i\,\left(\mathbf{X}^m_{p,q}\right)^T\, \\ 
\hbox{column $j\to k$ with $k$ such that} & \quad & \mathbf{s}^{k} = \mathbf{s}^j\,\mathbf{V}^{-1} \, , \\ 
\hbox{symbol $k\to j$ with $k$ such that} &  & \mathbf{s}^k=\mathbf{s}^j\,%
\mathbf{V}\, .
\end{array}
\label{colperm_CNOTstandard_to_original}
\end{equation}

\subsection{Local Clifford operations}

\label{localcliffordoperations}

Recall that our monomials $Z_{\alpha}X_{\beta}$ of Eq.~(\ref{basicmonomials}) decompose into  direct products 
$$
Z_{\alpha}X_{\beta}={\mathcal{Z}}^{c_1a_1}{\mathcal{X}}^{b_1}\otimes {\mathcal{Z}}^{c_2 a_2}{\mathcal{X}}^{b_2}\otimes\ldots \otimes 
{\mathcal{Z}}^{c_n a_n}{\mathcal{X}}^{b_n} 
$$

We consider a class of local Clifford operations $\mathbf{U}=\mathbf{U}_1\otimes\mathbf{U}_2\otimes \ldots \mathbf{U}_n$, 
with $\mathbf{U}_i$ acting locally on qudit $i$, that result in a map from $Z_{\alpha}X_{\beta}$ to another Pauli operator 
\begin{equation}
\mathbf{U}\left(Z_{\alpha}X_{\beta}\right)\mathbf{U}^\dagger=Z_{\alpha^{%
\prime }}X_{\beta^{\prime }} ={\mathcal{Z}}^{c_1m_1}{\mathcal{X}}%
^{\ell_1}\otimes {\mathcal{Z}}^{c_2 m_2}{\mathcal{X}}^{\ell_2}\otimes\ldots
\otimes {\mathcal{Z}}^{c_n m_n}{\mathcal{X}}^{\ell_n}
\end{equation}
where $\mathcal{Z}^{c_im_i}\mathcal{X}^{\ell_i}=\mathbf{U}_i\left(\mathcal{Z}%
^{c_ia_i}\mathcal{X}^{b_i}\right)\mathbf{U}_i^{\dagger}$.

The $p\times p$ matrix $\mathbf{U}_i$ must therefore induce on the exponents $a_i$ and $b_i$ a map $T$: 
\begin{equation}
\mathbf{U_i}\to T(\mathbf{U_i})=\left( 
\begin{array}{cc}
k^i_{11} & k^i_{12} \\ 
k^i_{21} & k^i_{22}%
\end{array}
\right), \qquad k^i_{st}\in \mathbb{Z}_{p}\, ,\quad \det[ \mathbf{U}_i]= 1 \, ,
\label{T_Ui}
\end{equation}
so that $\mathcal{Z}^{c_ia_i}\mathcal{X}^{b_i}$ goes to $\mathcal{Z}^{c_im_i}%
\mathcal{X}^{\ell_i}$, where 
\begin{equation}
\left(%
\begin{array}{c}
m_i \\ 
\ell_i%
\end{array}%
\right)=T(\mathbf{U}_i) \left(%
\begin{array}{c}
a_i \\ 
b_i%
\end{array}%
\right) \, .  \label{T_Ui_action}
\end{equation}

The effect of local Clifford operations has been investigated in \cite{beigi}, and we borrow their formalism. 
% to analyze the transformation of $\alpha(\sigma^i)$ and $\beta(\sigma^i)$.  
The analysis is simplified by noting that a set of commuting monomials,
augmented with the identity matrix, is an Abelian group of order $p^{n}$,
obtained from a set $G=\{g_{1},g_{2},\ldots g_{n}|g_{i}\in \mathbb{F}%
_{p^{n}}\}$ of $n$ generating elements. Conjugation by $\mathbf{U}$ simply
maps the original set $G$ of generating elements to another generating set $G^{\prime
}=\{g_{1}^{\prime },g_{2}^{\prime },\ldots g_{n}^{\prime }\}$.

We introduce, following \cite{beigi}, the $n\times 2n$ generator matrix $%
\mathbf{A}^{(f)}$, defined on generating elements $g_i$ by 
\begin{equation}
\mathbf{A}^{(f)}_{k,i}=\tr[\theta_k\,g_i]\, ,\quad \mathbf{A}^{(f)}_{k,i+n}=%
\tr[c_k^{-1}\theta_k\,f(g_i)]\, ,\qquad k,i=1,\ldots,n\, .  \label{A_matrix}
\end{equation}
The generator matrix $\mathbf{A}^{(f)}$ has the additional property that for
non-degenerate curves, it is a non-degenerate matrix, i.e. both  $n\times n$ submatrices  of the
matrix have non-zero determinants.

Choosing $G=\{g_{i}=c_{i}^{-1}\theta _{i},i=1,\ldots ,n\}$ such that the
monomials have the form $Z_{c_{i}^{-1}\theta _{i}}X_{f(c_{i}^{-1}\theta_{i})}$ 
reduces the matrix $\mathbf{A}^{(f)}$ to a suitably simple form: 
\begin{equation}
\mathbf{A}^{(f)}=(\,\mathinner{\hbox{1}\mkern-4mu\hbox{l}}\,|\,\mathbf{%
\Gamma }^{(f)}\,)\, ,
\end{equation}
while for a parametric curve $(\alpha (\sigma ^{i}),\beta (\sigma ^{i}))$, the generator matrix $\mathbf{A}^{(f)}$ 
is of the form 
\begin{equation}
\mathbf{A}^{(\alpha ,\beta )}=(\mathbf{\Gamma }^{(\alpha )}\mathbf{C}\,|\,%
\mathbf{\Gamma }^{(\beta )}).  \label{A_as_ga_gb}
\end{equation}%
It is then easy to verify that conjugation by $\mathbf{U}$ transforms $%
\mathbf{A}^{(f)}$ to 
\begin{eqnarray}
\mathbf{A}^{(f)}\rightarrow \mathbf{A}^{(\alpha ^{\prime },\beta ^{\prime})} 
&=&\left( \ \mathinner{\hbox{1}\mkern-4mu\hbox{l}}\ |\ \mathbf{\Gamma }^{(f)}\ \right) \left( 
\begin{array}{cc}
\mathbf{K}_{11} & \mathbf{K}_{12} \\ 
\mathbf{K}_{21} & \mathbf{K}_{22}
\end{array} \right)   \nonumber \\
&=&\left( \,\mathbf{K}_{11}+\mathbf{\Gamma }^{(f)}\mathbf{K}_{21}\,\,|\mathbf{K}_{12}+\mathbf{\Gamma }^{(f)}\mathbf{K}_{22}\right) ,  
\label{AL1}
\end{eqnarray}%
where the diagonal matrices 
\begin{equation}
\mathbf{K}_{s}=\left( 
\begin{array}{ccc}
k_{s}^{1} & 0\ldots  & 0 \\ 
0 & \ddots  & 0 \\ 
0 & \ldots 0 & k_{s}^{n}%
\end{array}%
\right) \,,\quad s=(11),(12),(21),(22)\,.
\end{equation}%
The new generator matrix has entries $\mathbf{A}^{(\alpha ^{\prime },\beta^{\prime })}$ resulting in a transformation of 
the curve $\beta =f(\alpha )$ to $(\alpha ^{\prime },\beta ^{\prime })$. The new curve is invertible if 
$\det [\mathbf{\Gamma }^{(\alpha ^{\prime })}]
\neq 0$ and $\det[\mathbf{\Gamma }^{(\beta ^{\prime })}]\neq 0$. 
The LS corresponding to these adjacency matrices is given as before in (\ref{DLSG}). Eq. (\ref{Gf_as_GaGb}), which 
describes the composition 
$f=\beta \circ \alpha ^{-1}$, can be used to bring $\mathbf{A}^{(\alpha ^{\prime },\beta ^{\prime })}$ to the form 
$(\mathinner{\hbox{1}\mkern-4mu\hbox{l}}\,|\,\mathbf{\Gamma }^{(f^{\prime })})$
from which the standard LS can be constructed using $f^{\prime }$.

\subsection{Composition of curves}\label{sec:compositionofcurves}

The adjacency matrix of a composed curve $\beta =f(g(\alpha ))$ is just
the product of the corresponding adjacency matrices as in Eq. (\ref{Gf_as_GaGb}).  Thus we
observe that LS transform under composition as 
\begin{equation}
L_{ij}^{f\circ g}=\left( \mathbf{s}^{j}\,\mathbf{C}^{-1}+\mathbf{s}^{i}\,%
\mathbf{\Gamma }^{(g)}\,\mathbf{C}\,\mathbf{\Gamma }^{(f)}\right) \bm{\theta}\, . \label{compos}
\end{equation}%
The transformation $\mathbf{s}^{i}\rightarrow \mathbf{s}^{i}\,\mathbf{\Gamma }^{(g)}\,\mathbf{C}$ is a column 
permutation, so the composition rule allows the construction of \emph{orbits} of LS in the same set of MOLS.
Such orbits are obtained by repeated composition of the type $\beta=f\circ f\circ \ldots \circ f(\alpha )$. In the particular 
case of the so-called Desarguesian bundle $\beta =\lambda \alpha $, the corresponding MOLS
contain  a single  orbit, generated from $f(\alpha )=\sigma \alpha $. In this way all LS in this set of MOLS can be 
obtained by (cyclic) permutations of  rows  in the LS corresponding to the $\beta =\sigma \alpha$  orbit.

\section{Operations on complete sets of MUBs and MOLS}

We are now in a position to discuss operations on a complete set of MUBs
described by the maximum number $p^{n}-1$ of distinct invertible curves.
These distinct curves produce a complete set of $p^{n}-1$ MOLS. Although an
individual curve admits a large number of unitaries that keep it invertible,
the situation is drastically different if we consider all the invertible
curves from a set of MUBs described by a bundle of curves corresponding to a
set of MOLS.

It is known that a complete set of MOLS (of dimension $d\times d$) exists if
and only if a finite projective plane of order $d$ also exists \cite{laywine_mullen, denes_keedwell1}. Of relevance to our
discussions are Desarguesian planes, which are based on linear equations
over finite fields, i.e. curves of the form $\{f_{\lambda}(\sigma^i)=\lambda\sigma^i; \lambda\in \mathbb{F}^{*}_{p^n}\}$ with $N=p^n$. 
A Desarguesian plane exists whenever $d=p^n$. Beyond their associations to
Desarguesian planes, the set (or bundle) of linear curves can also be used
to obtain monomials describing a complete set of MUBs. It is therefore
natural to speak of ``Desarguesian MUBs".

Additional planes, not of the Desarguesian type, can also exist; these
additional planes are not based on linear equations over finite fields.
Although there are MOLS associated with those planes, we have not found
examples of bundles of curves describing non-Desarguesian planes which
result in a complete set of MUBs of the monomial type. We should also point
out that there may be MUBs not of the monomial type which could be
associated with MOLS.

In this section we explore transformations of  one set of MUBs to another while preserving mutual
unbiasedness; we map one set of associated MOLS to another and find that such transformations correspond to an 
isomorphism of the 
corresponding MOLS. Two sets of MOLS are isomorphic iff there exists permutations of rows, columns, and symbols such 
that if applied to every 
square in the first set one obtains every square in the second set \cite{laywine_mullen}. Two sets of MOLS are also 
isomorphic if they are `built' on
the same affine (and consequently projective) plane \cite{denes_keedwell1}.

\subsection{CNOT for a complete set of invertible curves}

\label{subsec:CNOTforcompleteset}

A CNOT operation does not change the invertibility property of curves, as
per Eq.(\ref{CNOT_adjmat}). Suppose then, we have a complete set of MUBs
described by the maximum number $p^n - 1$ distinct invertible curves;
associated to this set (or bundle) of curves is a complete set of $p^n - 1$
MOLS. Applying the same CNOT transformation to each curve, we obtain another
set of invertible curves, associated to a different  but isomorphic set of
MOLS.

To see the action of a CNOT transformation on a complete set of curves, let
us choose as our initial bundle the rays associated to the MOLS in standard
form, 
\begin{equation}
f_{\lambda }\left( \alpha \right) =\lambda \alpha ,\qquad \lambda \in 
\mathbb{F}_{p^{n}}^{\ast }\, .  \label{desarguesiancurves}
\end{equation}
Thus
\begin{eqnarray}
g_{\lambda }\left( \alpha \right) &=&\hbox{CNOT}_{pq}\left[ f_{\lambda
}\left( \alpha \right) \right]  \nonumber \\
&=&\left(\alpha +\hbox{Tr}\left(\alpha\theta_q\right)\,c_p^{-1}\theta_p\right)\lambda\nonumber \\
&&+
\left(\hbox{Tr}\left(\lambda\alpha\,c_p^{-1}\theta_p\right)
+\hbox{Tr}\left(\alpha\theta_q\right)\hbox{Tr}\left(\lambda \,c_p^{-2}\theta^2_p\right)\right)\theta_q\, .
\end{eqnarray}

Since every invertible curve $f_{\lambda}$ transforms under the CNOT to some new
invertible curve $g_{\lambda}$, we can use Eq.~(\ref{CNOT_adjmat}) to obtain the
adjacency matrices of each $g_{\lambda}$ after application, and then Eq.~(\ref{OLSG})
to recover their Latin square in standard form.  This gives a new set of LS in standard form. An explicit example is provided
in \ref{subsec:3qubits}.

For each new Latin square so obtained, we can use Eqs.~(\ref{colperm_CNOTstandard_to_original}) to compute the 
permutations required to
take us back to the corresponding Desarguesian squares. These  permutations
are all identical, meaning that initial and transformed sets are indeed isomorphic.

The CNOT transformation given above yields isomorphic MOLS but changes the
separability structure of the associated MUBs. This structure is a common
means of classifying MUBs, as MUBs with different factorization structures
cannot be related by local unitary  transformations  \cite{romero, lawrence, JPAAndres}. Thus,
a phase-space approach where monomials are eventually associated with MOLS
provides structural information about the MUBs not contained in  their factorization structure. Alternatively, the 
factorization structure of the MUBs is not reflected at the level of phase-space.

In summary, even if there is a change in the factorization properties
resulting from application of a CNOT transformation, the associated MOLS
will remain isomorphic and will be isomorphic to the Desarguesian set.

\subsection{MOLS and local operations}

\label{MOLSandLocalOperations}

The transformation properties of \emph{individual} curves was discussed
using the generator matrix $\mathbf{A}$ of Section \ref%
{localcliffordoperations}. We now look at restrictions of these
transformations when they are applied to a \emph{bundle} of curves, using
again the generator matrix $\mathbf{A}$. A local unitary transformation is
determined by $4$ field elements $k_{s}$ through the map of Eq. (\ref%
{T_Ui_action}), so we can obtain necessary conditions by examining a limited
number of functions carefully chosen to fully constrain these parameters.

A bundle of invertible curves $f_i(\alpha), i = 1,\ldots, p^n-1$ correspond
to both a complete set of MUBs and MOLS. Since MUBs consist of disjoint sets
there will always be one and only one function in the bundle which, for $\sigma_k$ fixed but otherwise
arbitrary, maps some $\sigma_i$ to this $\sigma_k$. For some fixed $q$, there exists a function - let's
call it $F^{q}_m$ - with the property that
\begin{equation}
F^q_m (\theta_q) = m c_q \theta_q,\, \qquad m \in \mathbb{Z}^*_p.
\end{equation}
In other words, they map one (almost) self-dual basis element to a multiple of itself.
We will continue by limiting the discussion to $2$ particles;
this argument can of course be generalized to any number of particles.

The virtue of $F^q_m$ is that the matrix $\mathbf{A}^{(F^q_m)}$ takes a particularly 
useful form on the generating
elements $\{c_k^{-1}\theta_k\}$: the adjacency matrix will have in the $k^{\hbox{th}}$ row and the  $k^{\hbox{th}}$ column
a single diagonal entry with value $m$:
\begin{equation}
\left(\mathbf{\Gamma}^{(F^q_m)} \right)_{kq} = \hbox{Tr} \left[ c_k^{-1}
\theta_k F^q_m (c_q^{-1} \theta_q) \right] = m \delta_{kq} \, ,
\end{equation}
where $k$ and $q$ label the generating elements. 
Selecting $q=1$ for the purpose of the argument, and denoting entries unnecessary to our argument by $*$, we see that,
under a local unitary transformation $\mathbf{U_1}$ on the first particle, 
with $2\times 2$ representation $T(\mathbf{U}_1)$, $\mathbf{A}^{(F_{m}^{1})}$ is
transformed to 
\begin{eqnarray}
\mathbf{A}^{(F^{'1}_m)}&=& \left(
\begin{array}{cc|cc}
1 & 0 & m & 0 \\ 
0 & 1 & 0 & *%
\end{array}%
\right) \left(\begin{array}{cc:cc} k^1_{11}&0&k^1_{12}&0\\ 0&1&0&0\\
\hdashline k^1_{21}&0&k^1_{22}&0\\ 0&0&0&1\end{array}\right)\, ,  \nonumber
\\
&=& \left(
\begin{array}{cc|cc}
k_{11}^1+mk_{21}^1 & 0 & k_{12}^1+mk_{22}^1 & 0 \\ 
0 & 1 & 0 & *
\end{array}
\right)\, .
\end{eqnarray}

We are assuming det$[T(\mathbf{U}_1)]=k_{11}^1k_{22}^1-k_{12}^1k_{21}^1=1$. In order to preserve
non-degeneracy of the resulting curve, we must additionally have: 
\begin{equation}
k^1_{11} + m k^1_{12} \neq 0\, ,\quad\hbox{and}\quad k^1_{21} + m k^1_{22}
\neq 0.  \label{kconditions}
\end{equation}
There are thus only two possible forms for the matrix $T(\mathbf{U}_1)$.  One possibility is to suppose $k_{11}^1\ne 0$. 
Then, to guarantee the 
first of Eq.~(\ref{kconditions}) we must have $k_{21}^1=0$. To guarantee the condition on the determinant, we must now 
have $k_{22}^1\ne 0$, 
which in turn implies $k_{12}^1=0$ since Eq.~(\ref{kconditions}) must hold for arbitrary $m$. 
In fact, using the determinant condition we find
$k_{22}^1=(k_{11}^1)^{-1}$. One then obtains $\mathbf{K}$ as 
\begin{equation}
\left(
\begin{array}{cc}
\mathbf{K}_{11} & \mathbf{K}_{12} \\ 
\mathbf{K}_{21} & \mathbf{K}_{22}%
\end{array}%
\right)= \left( {\renewcommand{\arraystretch}{0.85} \renewcommand{%
\arraycolsep}{2.95pt} {\small \begin{array}{cc:cc} k^1_{11}&0&0&0\\
0&1&0&0\\ \hdashline 0&0&(k_{11}^1)^{-1}&0\\ 0&0&0&1\end{array} }} \right)=
\left(%
\begin{array}{cc}
\mathbf{K}_{11} & \mathbf{0} \\ 
\mathbf{0} & (\mathbf{K}_{11})^{-1}%
\end{array}
\right)
\end{equation}
%The matrix $T(\mathbf{U}_{j})$, A
Acting on qudit $j$, this is a \emph{scaling} transformation:
\begin{equation}
T(\mathbf{U}^S_{j})=\left( 
\begin{array}{cc}
k_{j} & 0 \\ 
0 & k_{j}^{-1}
\end{array}
\right)\, ,
\end{equation}
(S for scaling):  under $\mathbf{U}_j^{S}(k)$  the monomials transform as
\begin{equation}
\mathcal{Z}^{c_{1}a_{1}}\mathcal{X}^{b_{1}}\rightarrow \mathcal{Z}%
^{c_{1}k_ja_{1}}\mathcal{X}^{k_j^{-1}b_{1}}=(\mathcal{Z}^{c_{1}a_{1}})^{k_j}(%
\mathcal{X}^{b_{1}})^{k_j^{-1}}\,,
\end{equation}%
sending $\mathcal{Z}\rightarrow \mathcal{Z}^{k_j}$ and $\mathcal{X}\rightarrow \mathcal{X}^{k_j^{-1}}$.  
We call a matrix of the type 
\begin{equation}
\left( 
\begin{array}{cc}
\mathbf{K}_{11} & 0 \\ 
0 & (\mathbf{K}_{11})^{-1}%
\end{array}%
\right)  \label{typeitransformation}
\end{equation}%
a \emph{Type S transformation.}

For the second case, suppose instead that $k_{11}^1=0$. Necessarily $%
k_{21}^1\ne 0$ and $k_{12}^1\ne 0$ to preserve det$[T(\mathbf{U}_1)]=1$. But
we can choose to work with any $F^{q}_m$, meaning $m$ is arbitrary, which
implies in turn that $k_{22}^1=0$ to guarantee the second of Eq.~(\ref%
{kconditions}) always holds. The same conclusion is reached starting with
the assumption $k_{22}^1=0$.

We then obtain 
\begin{equation}
\left(%
\begin{array}{cc}
\mathbf{K}_{11} & \mathbf{K}_{12} \\ 
\mathbf{K}_{21} & \mathbf{K}_{22}%
\end{array}%
\right)= \left( {\renewcommand{\arraystretch}{0.85} \renewcommand{%
\arraycolsep}{2.95pt} {\small \begin{array}{cc:cc} 0&0&-k_j^{-1}&0\\
0&1&0&0\\ \hdashline k_j &0&0&0\\ 0&0&0&1\end{array} }} \right) = \left( 
\begin{array}{cc}
\mathbf{K}_{11} & -\overline{\mathbf{K}}_{21} \\ 
\mathbf{K}_{21} & \mathbf{K}_{11}%
\end{array}
\right)  \label{typeftransformation}
\end{equation}
where $\overline{\mathbf{K}}_s$ is the matrix with the inverses of the
non-zero elements of $\mathbf{K}_s$. Recalling 
\begin{equation}
T(\mathbf{U}_j) = \left( 
\begin{array}{cc}
0 & -k_j^{-1} \\ 
k_j & 0%
\end{array}
\right)
\end{equation}
we can see that the transformation on the monomials is now 
\begin{equation}
\mathcal{Z}^{c_1 a_1} \mathcal{X}^{b_1} \rightarrow \mathcal{Z}^{k_j b_1} \mathcal{X}^{- c_1
k_j^{-1} a_1}
\end{equation}
and the indices $a_1$ and $b_1$ are interchanged.

Consider the two qubit case, for which $c_{1}=1$ and $a_{i},b_{i}\in \mathbb{Z}_{2}$. 
If $a_{1}=0$ and $b_{1}=1$, $\mathcal{X}\rightarrow \mathcal{Z}$
and $\mathcal{Z}\rightarrow \mathcal{X}$. This means that in the qubit case,
this transformation collapses to a Fourier swap of $\mathcal{Z}$ and $%
\mathcal{X}$. Thus, we call a transformation having the form of Eq.~(\ref%
{typeftransformation}) a \emph{Type F transformation}.

These transformations have a form where they explicitly act on the $j$'th qudit: 
\begin{equation}
\mathbf{U}_j^{S}(r)=\sum_{t }|t_j \rangle \langle rt_j |,\quad \mathbf{U}_j^{F}(r)=\sum_{t ,m} \frac{\omega ^{rmt_j }}{\sqrt{d}}|t_j \rangle \langle m_j|.
\end{equation}

The conditions on the $k_{s}^{i}$ are necessary conditions for all curves in
the bundle: if they do not hold then at least some of the $F_{m}^{q}$
curves will not be invertible. We establish now that when applied on \emph{any}
curve, a combination of Type S and Type F transformations produces at least
one non-invertible curve.

Suppose, using our two-particle system, we apply a Type S transformation to
the first particle, and a Type F to the second. Then, for instance: 
\begin{equation}
T(\mathbf{U}_{1}^{S}(r))=\left( 
\begin{array}{cc}
r & 0 \\ 
0 & r^{-1}%
\end{array}%
\right) , \\
T(\mathbf{U}_{2}^{F}(t))=\left( 
\begin{array}{cc}
0 & -t^{-1} \\ 
t & 0%
\end{array}%
\right)
\end{equation}%
and 
\begin{equation}
\mathbf{K}=\left( \begin{array}{cc:cc} r & 0 & 0 & 0 \\ 0 & 0 & 0 & -t^{-1}
\\ \hdashline 0 & 0 & r^{-1} & 0\\ 0 & t & 0 & 0 \end{array}\right) =\left( 
\begin{array}{cc}
\mathbf{K}_{11} & -\overline{\mathbf{K}}_{21} \\ 
\mathbf{K}_{21} & \overline{\mathbf{K}}_{11}%
\end{array}%
\right) .
\end{equation}%
Applying this to $\mathbf{A}^{(f)}=(\mathinner{\hbox{1}\mkern-4mu\hbox{l}}|%
\mathbf{\Gamma }^{(f)})$, we obtain a transformed generator matrix 
\begin{eqnarray}
\mathbf{A}^{(f^{\prime })}=\mathbf{A}^{(f)}\mathbf{K} &=&\left( \mathbf{K}%
_{11}+\mathbf{\Gamma }^{(f)}\mathbf{K}_{21}|-\overline{\mathbf{K}}_{21}+%
\mathbf{\Gamma }^{(f)}\overline{\mathbf{K}}_{11}\right)  \nonumber \\
&=&\left( 
\begin{array}{cc|cc}
r & \Gamma _{12}^{(f)}t & \Gamma _{11}^{(f)}r^{-1} & 0 \\ 
0 & \Gamma _{22}^{(f)}t & \Gamma _{21}^{(f)}r^{-1} & -t^{-1}%
\end{array}%
\right) .
\end{eqnarray}

In a bundle of $p^n-1$ curves, there will always be one curve such that $%
f(\theta_1) \propto \theta_2$ (otherwise it would not be complete). For this
curve, $\Gamma^{(f)}_{11} = 0$, and thus the second half of the transformed
generator matrix will be degenerate, meaning this curve is no longer
invertible. This argument can be generalized to any number of particles.
Thus, we are limited to local transformations where \emph{all}
transformations on the particles are either Type S, or all Type F.

Then: 
\begin{equation}
\mathbf{A}_{S}^{(f^{\prime })}=\left( \,\mathbf{K}_{11}\,|\mathbf{\Gamma }^{(f)} 
\mathbf{K}_{11}^{-1}\right) ,\quad \mathbf{A}_{F}^{(f^{\prime })}=\left( \,%
\mathbf{\Gamma }^{(f)}\mathbf{K}_{21}|-\mathbf{K}_{21}^{-1}\right) ,
\end{equation}
where $\mathbf{A}_{S}^{(f^{\prime })}$ and $\mathbf{A}_{F}^{(f^{\prime })}$
are matrices after transformations of types S and F, respectively.

Following Eq.(\ref{A_as_ga_gb}),  these transformations correspond to the new  parametric 
curves 
\begin{equation}
\mathbf{\Gamma }^{( \alpha')}=\mathbf{K}_{11}\mathbf{C}^{-1},\quad \mathbf{\Gamma }^{( \beta')} =
\mathbf{\Gamma }^{(f)}\mathbf{K}_{11}^{-1}, \quad \hbox{for  S-type}\, ,
\end{equation}
and 
\begin{equation}
\mathbf{\Gamma }^{( \alpha')}=\mathbf{\Gamma }^{(f)}\mathbf{K}_{21}\mathbf{C}^{-1},\quad 
\mathbf{\Gamma }^{(\beta')}=-\mathbf{K}_{21}^{-1}, \quad \hbox{for  F-type}.
\end{equation}

The corresponding LS are obtained from the originals by permutations. Observing that $\mathbf{C}$
and $\mathbf{K}_{ij}$ commute, we must apply the 
permutations
\begin{equation}
\begin{array}{rcl}
\hbox{no permutation of rows,} &  & \\
\hbox{column } i\to k \hbox{ with $k$ such that} & \quad & \mathbf{s}^{k}=%
\mathbf{s}^{i}\,\mathbf{K}_{11}^{2}\, , \\ 
\hbox{symbol } k \to j \hbox { with $k$ such that } & \quad & \mathbf{s}^k = \mathbf{s}^j \mathbf{K}_{11}^{-1}
\end{array}
\end{equation}
for Type S transformations.

For Type F transformations, we start from
\beq
\tilde L^{(f')}_{ij}=\left(\mathbf{s}^j\bm{\Gamma}^{(f)}\mathbf{K}_{21}\mathbf{C}^{-1}-\mathbf{s}^i\mathbf{K}_{21}^{-1}\right)\bm{\theta}
\eeq
and we first interchange the rows and columns.   
Then, to this transformed square, we can 
apply the permutations
\begin{equation}
\begin{array}{rcl}
\hbox{no permutation of rows,} &  & \\
\hbox{column } j\to k \hbox{ with $k$ such that} & \quad & \mathbf{s}^{k}= - \mathbf{s}^{j}\,\mathbf{K}_{21}^{-2}\, 
\mathbf{C}^2\, , \\ 
\hbox{symbol } k \to j \hbox { with $k$ such that } & \quad & \mathbf{s}^k = 
\mathbf{s}^j \,\mathbf{C}^{-1}\, \mathbf{K}_{21}.
\end{array}
\end{equation}
Although transposition is not an isotopy, the transformed and original squares are still main class equivalent.  
We note the interesting coincidence that 
a Fourier transformation will interchange two complementary variables, corresponding to an interchange in the phase-space axes, much like
the transposition needed to bring $\tilde L^{(f')}_{ij}$ back to $L^{(f)}_{ij}$.

This set of local transformations has interesting composition relations
with CNOT operations. If $\mathbf{U}^{S}$ and $\mathbf{U}^{F}$ are
transformations  of type S and F respectively, 
\begin{eqnarray}
\mathbf{U}_{p}^{S}(r)\mathbf{U}_{q}^{S}(t)\mathbf{X}_{p,q}^{m} &\sim &%
\mathbf{X}_{p,q}^{mtr^{-1}}\mathbf{U}_{p}^{S}(r)\mathbf{U}_{q}^{S}(t) \nonumber \\
\mathbf{U}_{p}^{F}(r)\mathbf{U}_{q}^{F}(t)\mathbf{X}_{p,q}^{m} &\sim &%
\mathbf{X}_{p,q}^{-mtr^{-1}}\mathbf{U}_{p}^{F}(r)\mathbf{U}_{q}^{F}(t)  \label{compositionrelation}
\end{eqnarray}%
where $\sim $ indicates these hold to within an overall phase. On the other
hand, the composition and commutation relations between $\mathbf{U}_{j}^{F}$
and $\mathbf{U}_{j}^{S}$ are as follows:%
\begin{eqnarray}
\mathbf{U}_{j}^{S}(k)\mathbf{U}_{j}^{S}(r) &=&\mathbf{U}_{j}^{S}(kr),\quad 
\mathbf{U}_{j}^{F}(k)\mathbf{U}_{j}^{F}(r)=\mathbf{U}_{j}^{S}(-kr^{-1}),  \label{LP1} \\
\mathbf{U}_{j}^{S}(k)\mathbf{U}_{j}^{F}(t) &=&\mathbf{U}_{j}^{F}(tk^{-1}),%
\quad \mathbf{U}_{j}^{F}(t)\mathbf{U}_{j}^{S}(k)=\mathbf{U}_{j}^{F}(tk).
\label{LP2}
\end{eqnarray}

We conclude that unitary transformations on MUBs preserving the complete
set of MOLS must be a combination of CNOT transformations, or Type S/F
transformations, where either Type S transformations are applied to all
qudits, or Type F to all qudits.

This conclusion is important as it allows us to drastically simplify the
possible sequence of transformations that map invertible curves to
invertible curves. As a result, all non-local transformations can be done
consecutively, followed by a sequence of local transformations.

\section{Latin minisquares and commutative curves}

\label{sec:minisquares}

Curves corresponding to Latin squares are permutation polynomials \cite{permutationpolynomials}; these curves are 
necessarily invertible,  but may not be commutative. However, only curves which are both additive and
commutative can correspond to MUBs of monomial type \cite{curvepapers}.
Given an arbitrary Latin square, we provide a method of quickly determining if its associated curve is commutative.

We define a minisquare $\ell^{(f)}$, a subsquare of its parent $L^{(f)}$, with entries computed from only the self-dual basis elements, 
\begin{equation}
\ell_{ij}^{(f)}=\theta _{j}+f\left(c^{-1}_i \theta _{i}\right) ,\quad
i,j=1,...,n.  \label{minisquare}
\end{equation}

Writing $\theta_j=\sigma^{p(j)}$ and $c_i^{-1}\theta_i=\sigma^{q(i)}$ for
some functions $p$ and $q$, we have 
\begin{eqnarray}
&\ell_{ij}^{(f)}&=L^{(f)}_{ q(i)p(j) },  \label{ms_ls_relationship} \\
\hbox{Tr}\left( \ell^{(f)}_{ij}c_{j}^{-1}\theta _{j}\right) &=&1+\Gamma
_{ji}^{(f)}\, .  \nonumber
\end{eqnarray}
The minisquares thus contain information about $\Gamma _{ji}^{(f)}$. In
addition, $\ell^{(f)}$ corresponds to a commutative curve if 
\begin{equation}
\hbox{Tr}\left( \ell_{ij}c_{j}^{-1}\theta _{j}\right) =\hbox{Tr}\left(
\ell_{ji}c_{i}^{-1}\theta _{i}\right) .
\end{equation}

One example of application is provided in \ref{subsec:3qubits}. Here, we
consider instead a situation where LS from a set associated to the Hall
projective plane in dimension $9$ do not have associated MUBs of the monomial type. This is a 2-qutrit problem; since the
(almost) self-dual basis is $\{ \sigma^4, \sigma^2\}$ we find $p(1) = 4$, $%
p(2) = 2$, $q(1) = 8$ and $q(2) = 2$ since $c_1=2$.

We start with a LS obtained from the Hall plane, with associated curve $f =
\sigma^5 \alpha^3$: 
\begin{equation}
L^{(\sigma^5 \alpha^3)} = \left( {\renewcommand{\arraystretch}{0.85} %
\renewcommand{\arraycolsep}{2.95pt} {\small 
\begin{array}{ccccccccc}
0 & 1 & 2 & 3 & 4 & 5 & 6 & 7 & 8 \\ 
8 & 7 & 3 & 5 & 0 & 2 & 1 & 6 & 4 \\ 
3 & 4 & \bm{1} & 7 & \bm{2} & 6 & 8 & 0 & 5 \\ 
6 & 3 & 0 & 8 & 7 & 4 & 2 & 5 & 1 \\ 
1 & 5 & 8 & 4 & 6 & 0 & 3 & 2 & 7 \\ 
4 & 6 & 5 & 2 & 8 & 3 & 7 & 1 & 0 \\ 
7 & 2 & 4 & 0 & 1 & 8 & 5 & 3 & 6 \\ 
2 & 8 & 6 & 1 & 5 & 7 & 0 & 4 & 3 \\ 
5 & 0 & \bm{7} & 6 & \bm{3} & 1 & 4 & 8 & 2%
\end{array}
}} \right).
\end{equation}

The elements of the minisquare are then 
\begin{equation}
\ell^{( \sigma^5 \alpha^3)} = \left( 
\begin{array}{cc}
L_{84} & L_{82} \\ 
L_{24} & L_{22}%
\end{array}
\right) \to \left( 
\begin{array}{cc}
\sigma^3 & \sigma^7 \\ 
\sigma^2 & \sigma%
\end{array}
\right) = \left( 
\begin{array}{cc}
2 \theta_1 + \theta_2 & \theta_1 + 2 \theta_2 \\ 
\theta_2 & 2 \theta_1 + 2 \theta_2%
\end{array}
\right)\, .
\end{equation}

The curve associated with this LS is not commutative, since $\hbox{Tr}(\ell_{ij}c_j^{-1}\theta_j) \neq \hbox{Tr}(\ell_{ji} c_i^{-1} \theta_i)$ for $i \neq j$.

% The curve associated with this LS is not commutative, since 
% \begin{equation}
% \hbox{Tr}(\ell_{ij} c_j^{-1} \theta_j) = \left( 
% \begin{array}{cc}
% 2 & 2 \\ 
% 0 & 2%
% \end{array}
% \right) \quad\ne \quad \hbox{Tr}(\ell_{ji} c_i^{-1} \theta_i) = \left( 
% \begin{array}{cc}
% 2 & 0 \\ 
% 2 & 2%
% \end{array}
% \right)\, .
% \end{equation}
The resulting adjacency matrix, though invertible as required,
is not symmetric: 
\begin{equation}
\mathbf{\Gamma}^{(f)} = \left( 
\begin{array}{cc}
1 & 1 \\ 
2 & 1%
\end{array}
\right)\, 
\end{equation}
and so cannot describe a commuting curve.

\section{Conclusions}

We have shown that, in dimension $p^{n}$,  one can  associate a
complete sets of MUBs with a set of $p^{n}-1$ of invertible curves, and thus
with a complete set of MOLS (excluding two``faux'' squares obtained from degenerate curves). There
exist subsets of unitary transformations acting on the MUBs that induce isomorphisms at the level of the corresponding 
MOLS. These transformations
comprise CNOT-type transformations, and local transformations denoted Type S and Type F applied uniformly to all
 the particles. This analysis has
also allowed us to unravel the isomorphism permutations between the MOLS associated with the original and transformed
 MUBs.

In particular, we explicitly provide the mappings between MUBs of the
monomial type described by invertible curves and their associated MOLS so
that,  given these restrictions, the following diagram holds:
$$
\begin{array}{ccc}
MUBs & \Leftrightarrow  & MOLS \\ 
\hbox{Unitary transformation}\Updownarrow  &  & \Updownarrow %
\hbox{Isomorphism permutation} \\ 
MUBs^{\prime } & \Leftrightarrow  & MOLS^{\prime }%
\end{array}%
$$

We have also shown that local transformations preserving mappings from MUBs
to MOLS form a group with multiplication given in Eqs. (\ref{LP1})-(\ref{LP2}).
The composition relation with CNOT given in Eq.~(\ref{compositionrelation}) always preserves the mapping.  
This relation allows us to separate permutations of LS  elements of  a given set of MOLS related by local and non-local
transformations, thus simplifying the classification of permutations that
preserve the relations between MOLS and MUBs.

Arbitrary Latin squares do not necessarily correspond to
MUBs, as is evident in the example of the Hall square of section 7.  
Latin minisquares serve as an excellent means of verifying  commutativity of a curve. 
Conversely, not every set of MUBs leads to a  set of MOLS, but only those that
contain $p^{n}-1$ invertible curves. Furthermore, arbitrary permutations on
Latin squares that do correspond to MUBs do not necessarily preserve them,
unless they conform to the types of transformations listed above.

\section{Acknowledgments}

We thank Prof. K. Hicks for his help in numerically testing
some of the hypotheses in the early stages of this work. ODM was funded in
part by NSERC of Canada. Part of this work was done while visiting the
Fields Institute in Toronto, and this visit was supported in part by the
Julian Schwinger Foundation. The work of ABK is supported by the Grant
106525 of CONACyT (Mexico). HdG acknowledges support from NSERC. ABK and
HdG also acknowledge partial support from the Fields Institute.

\appendix

\section{Two examples}

\subsection{MOLS and CNOT for 3 qubits}

\label{subsec:3qubits}

The irreducible polynomial is $\sigma ^{3}+\sigma ^{2}+1=0$; 
$\bm{\theta}=\{\sigma,\sigma^2,\sigma^4\}$, $\mathbf{C}=
\mathinner{\hbox{1}\mkern-4mu\hbox{l}}$.

The expansions in the self-dual basis are 
\begin{equation}
\begin{array}{llll}
\mathbf{s}^{0}=\left( 0,0,0 \right) & \mathbf{s}^{1}=\left( 1, 0, 0 \right)
& \mathbf{s}^{2}=\left(0, 1,0\right) & \mathbf{s}^{3}=\left( 1, 0,1 \right)
\\ 
&  &  &  \\ 
\mathbf{s}^{4}=\left( 0, 0,1\right) & \mathbf{s}^{5}=\left( 0,1, 1\right) & 
\mathbf{s}^{6}=\left( 1,1,0 \right) & \mathbf{s}^{7}=\left( 1,1,1 \right)%
\end{array}%
\end{equation}

We choose the ray $f(\alpha)=\alpha$ so $f=$Id and $\mathbf{\Gamma }%
^{(\alpha)}=\mathinner{\hbox{1}\mkern-4mu\hbox{l}}$. Using (\ref{OLSG}): 
\begin{equation}
L^{(\alpha)}= 
\begin{array}{c}
\\ 
\\ 
\\ 
\\ 
\\ 
\\ 
\\ 
\end{array}
\left( {\renewcommand{\arraystretch}{0.85} \renewcommand{\arraycolsep}{2.95pt} {\small 
\begin{array}{cccccccc}
0 & 1 & 2 & 3 & 4 & 5 & 6 & 7 \\ 
1 & 0 & 6 & 4 & 3 & 7 & 2 & 5 \\ 
2 & 6 & 0 & 7 & 5 & 4 & 1 & 3 \\ 
3 & 4 & 7 & 0 & 1 & 6 & 5 & 2 \\ 
4 & 3 & 5 & 1 & 0 & 2 & 7 & 6 \\ 
5 & 7 & 4 & 6 & 2 & 0 & 3 & 1 \\ 
6 & 2 & 1 & 5 & 7 & 3 & 0 & 4 \\ 
7 & 5 & 3 & 2 & 6 & 1 & 4 & 0%
\end{array}%
}} \right)  \label{OL 3qb}
\end{equation}
This Latin square is symmetric, as is always the case for the curve $\beta = \alpha$. A Latin square with both the first row and first column in standard
order is said to be \emph{reduced} \cite{laywine_mullen}.

Suppose we perform CNOT on the first and second qubits, $\mathbf{X}^1_{1, 2}$. By (\ref{CNOTm}) and (\ref{CNOT_adjmat}): 
\begin{equation}
\mathbf{X}^1_{1, 2} =\left( {\renewcommand{\arraystretch}{0.85} %
\renewcommand{\arraycolsep}{2.95pt} {\small 
\begin{array}{ccc}
1 & 1 & 0 \\ 
0 & 1 & 0 \\ 
0 & 0 & 1%
\end{array}
}} \right) \, ,\qquad \mathbf{\Gamma }^{(g) }= (\mathbf{X}^1_{1, 2} \,)^{T}%
\mathbf{\Gamma}^{(\alpha)} \mathbf{X}^1_{1,2} =\left( {\renewcommand{%
\arraystretch}{0.85} \renewcommand{\arraycolsep}{2.95pt} {\small 
\begin{array}{ccc}
1 & 1 & 0 \\ 
1 & 0 & 0 \\ 
0 & 0 & 1%
\end{array}
}} \right)  \label{Gammag}
\end{equation}
where $\mathbf{\Gamma}^{(\alpha)}=\mathinner{\hbox{1}\mkern-4mu\hbox{l}}$,
as noted before. The resulting square is: 
\begin{equation}
\tilde{L}^{(g) } = 
\begin{array}{c}
\\ 
\\ 
\\ 
\\ 
\\ 
\\ 
\\ 
\end{array}
\left( {\renewcommand{\arraystretch}{0.85} \renewcommand{%
\arraycolsep}{2.95pt} {\small 
\begin{array}{cccccccc}
0 & 1 & 6 & 3 & 4 & 7 & 2 & 5 \\ 
6 & 2 & 0 & 5 & 7 & 4 & 1 & 3 \\ 
2 & 6 & 1 & 7 & 5 & 3 & 0 & 4 \\ 
7 & 5 & 4 & 2 & 6 & 0 & 3 & 1 \\ 
4 & 3 & 7 & 1 & 0 & 6 & 5 & 2 \\ 
5 & 7 & 3 & 6 & 2 & 1 & 4 & 0 \\ 
1 & 0 & 2 & 4 & 3 & 5 & 6 & 7 \\ 
3 & 4 & 5 & 0 & 1 & 2 & 7 & 6%
\end{array}
}} \right)  \label{LSg_nonstandard}
\end{equation}

Using Eq.~(\ref{colperm_CNOT_nstos}), we can permute this square into
standard form. These equations produce $\mathbf{s}^2 \leftrightarrow \mathbf{%
s}^6 $ and $\mathbf{s}^5 \leftrightarrow \mathbf{s}^7 $, indicating that
rows 2 and 6, and rows 5 and 7 need to be interchanged, and likewise for the
columns. This yields 
\begin{equation}
L^{(g) }= 
\begin{array}{c}
\\ 
\\ 
\\ 
\\ 
\\ 
\\ 
\\ 
\end{array}
\left( {\renewcommand{\arraystretch}{0.85} \renewcommand{%
\arraycolsep}{2.95pt} {\small 
\begin{array}{cccccccc}
0 & 1 & 2 & 3 & 4 & 5 & 6 & 7 \\ 
6 & 2 & 1 & 5 & 7 & 3 & 0 & 4 \\ 
1 & 0 & 6 & 4 & 3 & 7 & 2 & 5 \\ 
7 & 5 & 3 & 2 & 6 & 1 & 4 & 0 \\ 
4 & 3 & 5 & 1 & 0 & 2 & 7 & 6 \\ 
3 & 4 & 7 & 0 & 1 & 6 & 5 & 2 \\ 
2 & 6 & 0 & 7 & 5 & 4 & 1 & 3 \\ 
5 & 7 & 4 & 6 & 2 & 0 & 3 & 1%
\end{array}
}} \right)  \label{XOL 3qb}
\end{equation}
This standard form square can also be found directly using $\mathbf{\Gamma}^{(g)}$ in Eq.~(\ref{Gammag}) and Eq.~(\ref{OLSG}).

To bring $L^{(g)}$ to $L^{(\alpha)}$, use Eq.~(\ref%
{colperm_CNOTstandard_to_original}). Since $\mathbf{C} = \mathinner{\hbox{1}%
\mkern-4mu\hbox{l}}$, the symbol permutation is obtained from 
\begin{equation}
(s_1^i, s_2^i ,s_3^i) \left( {\renewcommand{\arraystretch}{0.85} %
\renewcommand{\arraycolsep}{2.95pt} {\small 
\begin{array}{ccc}
1 & 1 & 0 \\ 
0 & 1 & 0 \\ 
0 & 0 & 1%
\end{array}
}} \right) \left( {\renewcommand{\arraystretch}{0.85} \renewcommand{%
\arraycolsep}{2.95pt} {\small 
\begin{array}{c}
\theta_1 \\ 
\theta_2 \\ 
\theta_3%
\end{array}
}} \right)\to \sigma^i\, ,  \label{example1symbolperm}
\end{equation}
so that, choosing $i=7$ for instance: 
\begin{equation}
(1,1,1) \left( {\renewcommand{\arraystretch}{0.85} \renewcommand{%
\arraycolsep}{2.95pt} {\small 
\begin{array}{ccc}
1 & 1 & 0 \\ 
0 & 1 & 0 \\ 
0 & 0 & 1%
\end{array}
}} \right) \left( {\renewcommand{\arraystretch}{0.85} \renewcommand{%
\arraycolsep}{2.95pt} {\small 
\begin{array}{c}
\sigma \\ 
\sigma^2 \\ 
\sigma^4%
\end{array}
}} \right) =\sigma+\sigma^4=\sigma^3\to \sigma^7\, .
\end{equation}
Proceeding systematically in this way we find the additional symbol
permutations $\sigma\leftrightarrow \sigma^6\, , \sigma^7\to \sigma^3$ with
all other unchanged. In the same manner, using again Eq.~(\ref%
{colperm_CNOTstandard_to_original}) it is found that the row transformations
are simply $2\leftrightarrow 6$ and $5\leftrightarrow 7$; the column
permutations are $1\leftrightarrow 6$ and $3\leftrightarrow 7$.

Alternatively, $\tilde{L}^{(g)}$ can be transformed back to $L^{(f)}$ of Eq.
(\ref{OL 3qb}) in a single shot, using Eqs. (\ref{colperm_CNOT_to_original})
as a starting point. 
%to transform the non-standard square (\ref{LSg_nonstandard}) into (\ref{OL 3qb}). 
%The self-dual basis is $\{\sigma,\sigma^2,\sigma^4\}$, and $\mathbf{C}=\unit$. 
No row permutation is necessary. The matrix 
\begin{equation}
\mathbf{W}=\left( {\renewcommand{\arraystretch}{0.85}\renewcommand{%
\arraycolsep}{2.95pt}{\small 
\begin{array}{ccc}
1 & 1 & 0 \\ 
1 & 0 & 0 \\ 
0 & 0 & 1%
\end{array}%
}}\right)
\end{equation}%
induces the column permutations % \beq
%  (\mathbf{s}^1)^T \rightarrow (\mathbf{s}^{6})^T \rightarrow (\mathbf{s}^{2})^T \rightarrow (\mathbf{s}^1)^T ;
%  \quad  (\mathbf{s}^3)^T \rightarrow (\mathbf{s}^{7})^T \rightarrow (\mathbf{s}^{5})^T \rightarrow (\mathbf{s}^3)^T\, .
% \eeq
\begin{equation}
1\rightarrow 6\rightarrow 2\rightarrow 1\qquad 
3\rightarrow 7\rightarrow 5\rightarrow 3
\end{equation}%
Finally, $\mathbf{X}_{1,2}^{1}$ induces the same symbol transformation $%
\sigma \leftrightarrow \sigma ^{6}\,,\sigma ^{3}\leftrightarrow \sigma ^{7}$
as before.

To illustrate the composition of curves of Sec.\ref{sec:compositionofcurves}, and the idea of orbits introduced in this section, 
we consider the adjacency matrix corresponding to $f(\alpha )=\sigma \alpha $:
\beq
\mathbf{\Gamma }^{(f)}=\left( 
\begin{array}{ccc}
0 & 1 & 0 \\ 
1 & 0 & 1 \\ 
0 & 1 & 1%
\end{array}%
\right) , 
\eeq
so that, using Eq.~(\ref{compos}) with $g=f$ leads to the cyclic
permutation of columns
\beq
7\rightarrow 6\rightarrow 5\rightarrow 4\rightarrow 3\rightarrow
2\rightarrow 1\rightarrow 7,
\eeq
which is nothing but the permutation of columns needed to take the LS corresponding to $f\circ f(\alpha)=\sigma^2 \alpha$ to back to $L^{(f)}$.

For completeness we provide in Table \ref{table:longtable} the complete set
of Desarguesian curves transformed under $\mathbf{X}_{1,2}^{1}$. The first
column shows the original curves whereas the second column show the
transformed curve. The new curves are calculated according to a
transformation known from \cite{graphstates}, valid for qubit curves: 
\begin{equation}
g(\alpha) = f(\alpha) + \hbox{Tr}(\alpha \theta_q ) f(\theta_p) + \hbox{Tr}
\lbrack f(\alpha) \theta_p] \theta_q + \hbox{Tr}(\alpha \theta_q) \hbox{Tr}%
\lbrack f(\theta_p) \theta_p] \theta_q
\end{equation}

\begin{table}[tbp]
\begin{center}
\begin{tabular}[H]{c||c}
initial curve & new curve \\ 
$f = \alpha$ & $g = \sigma^6 \alpha + \sigma \alpha^2 + \sigma^4 \alpha^4$
\\ \hline
$f = \sigma \alpha$ & $g = \sigma \alpha$ \\ \hline
$f = \sigma^2 \alpha$ & $g = \sigma^5 \alpha + \sigma^3 \alpha^2 + \sigma^5
\alpha^4$ \\ \hline
$f = \sigma^3 \alpha$ & $g = \sigma^3 \alpha + \sigma^4 \alpha^2 + \sigma^2
\alpha^4 $ \\ \hline
$f = \sigma^4 \alpha$ & $g = \sigma^4 \alpha + \sigma^4 \alpha^2 + \sigma^2
\alpha^4 $ \\ \hline
$f = \sigma^5 \alpha$ & $g = \sigma^2 \alpha + \sigma \alpha^2 + \sigma^4
\alpha^4 $ \\ \hline
$f = \sigma^6 \alpha$ & $g = \alpha + \sigma^3 \alpha^2 + \sigma^5 \alpha^4 $
\\ \hline
\end{tabular}
\label{table:longtable}
\end{center}
\caption{Two sets of curves, each corresponding to a different set of MUBs
with different separability properties. Application of the same sequence of
permutations to the MOLS obtained using the $f$ curves yields MOLS obtained
using the $g$ curves.}
\end{table}

To obtain the minisquare for $f(\alpha)=\sigma \alpha$ we first recall that 
$\bm{\theta}=\{\sigma,\sigma^2,\sigma^4\}$.
Since $\mathbf{C}=\mathinner{\hbox{1}\mkern-4mu\hbox{l}}$, $p(i)=q(i)$ and we find: 
\beq
p(1)=q(1)=1\, ,\qquad p(2)=q(2)=2\, ,\qquad 
p(3)=q(3)=4\, .
\eeq 
The minisquare then has entries which are those at the
intersections of lines and columns 1, 2 and 4 of (\ref{OL 3qb}) (recall the indexing starts with $0$): 
\begin{equation}
\ell ^{(\alpha )}=\left( {\renewcommand{\arraystretch}{0.85}%
\renewcommand{\arraycolsep}{2.95pt}{\small 
\begin{array}{ccc}
0 & 6 & 3 \\ 
6 & 0 & 5 \\ 
3 & 5 & 0%
\end{array}%
}}\right) \ \rightarrow \ \left( {\renewcommand{\arraystretch}{0.85}%
\renewcommand{\arraycolsep}{2.95pt}{\small 
\begin{array}{ccc}
0 & \sigma ^{6} & \sigma ^{3} \\ 
\sigma ^{6} & 0 & \sigma ^{5} \\ 
\sigma ^{3} & \sigma ^{5} & 0%
\end{array}%
}}\right) =\left( {\renewcommand{\arraystretch}{0.85}\renewcommand{%
\arraycolsep}{2.95pt}{\small 
\begin{array}{ccc}
0 & \theta _{1}+\theta _{2} & \theta _{1}+\theta _{3} \\ 
\theta _{1}+\theta _{2} & 0 & \theta _{2}+\theta _{3} \\ 
\theta _{1}+\theta _{3} & \theta _{2}+\theta _{3} & 0%
\end{array}%
}}\right)  \label{mLSe}
\end{equation}

The matrix with entries $\hbox{Tr}(\ell_{ij} \theta_j)$ is symmetric and
corresponds to $\Gamma^{(\alpha)}$ given by 
\begin{equation}
\hbox{Tr}\left( \ell_{ij} \theta_j\right) = \left( {\renewcommand{%
\arraystretch}{0.85} \renewcommand{\arraycolsep}{2.95pt} {\small 
\begin{array}{ccc}
0 & 1 & 1 \\ 
1 & 0 & 1 \\ 
1 & 1 & 0%
\end{array}
}} \right) \, ,\qquad \Rightarrow\qquad \Gamma^{(\alpha)}=%
\mathinner{\hbox{1}\mkern-4mu\hbox{l}}\, ,
\end{equation}
which, being symmetric, means $f$ is indeed commutative.

Finally, we mentioned in Section \ref{subsec:CNOTforcompleteset} that the
CNOT changes the separability properties of MUBs. In this example, the
eigenstates of MUB operators constructed from the linear functions $f$ have
separability structure (3, 0, 6), meaning 3 fully separable sets of states,
6 non-separable sets of states and one biseparable set of states. Eigenstates of
the transformed operators constructed from the functions $g$ have a different
structure: (2, 3, 4) (2 separable, 3 biseparable and 4 non-separable) \cite{romero}. Both sets however, are Desarguesian, illustrating how the
factorization structure is not reflected at the geometrical level.

\subsection{2 qutrits and a local transformation}

\label{subsec:2qutrits}

The irreducible polynomial is $\sigma ^{2}+\sigma +2=0$; the almost
self-dual basis $\bm{\theta}=\{\sigma^4, \sigma^2 \}$ produces 
\begin{equation}
\mathbf{C}=\left(%
\begin{array}{cc}
2 & 0 \\ 
0 & 1%
\end{array}%
\right)\, .
\end{equation}
Thus, $\sigma^i=2 s_1^i\sigma^4+s_2^i\sigma^2$. Explicitly, we have the
vectors: 
\begin{equation}
\begin{array}{ccccc}
\mathbf{s}^{0}=\left( 0,0 \right) & \mathbf{s}^{1}=\left( 1,2\right) & 
\mathbf{s}^{2}=\left( 0,1 \right) & \mathbf{s}^{3}=\left( 1,1\right) & 
\mathbf{s}^{4}=\left( 2, 0\right) \\ 
&  &  &  &  \\ 
& \mathbf{s}^{5}=\left( 2 ,1\right) & \mathbf{s}^{6}=\left( 0, 2\right) & 
\mathbf{s}^{7}=\left( 2, 2\right) & \mathbf{s}^{8}=\left( 1,0\right)\, .%
\end{array}%
\end{equation}

We choose the curve $\beta =\sigma ^{3}\alpha $, which can be represented in
parametric form %$Z_{\alpha(\sigma^i) }X_{\beta(\sigma^i) }$ with 
$(\alpha,\beta)$ with $\alpha \left( \sigma^i \right) =\sigma ^{2}\sigma^i $
and $\beta \left( \sigma^i \right) =\sigma ^{5}\sigma^i $, so %that
\begin{equation}
\mathbf{\Gamma }^{ (\alpha)}=\left( 
\begin{array}{cc}
0 & 1 \\ 
1 & 0%
\end{array}
\right) \, ,\qquad \mathbf{\Gamma }^{ (\beta)}=\left( 
\begin{array}{cc}
1 & 1 \\ 
1 & 2%
\end{array}%
\right).
\end{equation}
The points are not in standard order; the corresponding non-standard LS is
given by 
\begin{equation}
\tilde{L}^{(\sigma^3 \alpha)}_{ij}=\left[ \mathbf{s}^j\left( 
\begin{array}{cc}
0 & 1 \\ 
1 & 0%
\end{array}%
\right) +\mathbf{s}^i\left( 
\begin{array}{cc}
1 & 1 \\ 
1 & 2%
\end{array}%
\right) \right] \bm{\theta },
\end{equation}
which upon evaluation, gives the full square 
\begin{equation}
\tilde{L}^{(\sigma^3 \alpha)}= 
\begin{array}{c}
\\ 
\\ 
\\ 
\\ 
\\ 
\\ 
\\ 
\end{array}
\left( {\renewcommand{\arraystretch}{0.85} \renewcommand{%
\arraycolsep}{2.95pt} {\small 
\begin{array}{ccccccccc}
0 & 3 & 4 & 5 & 6 & 7 & 8 & 1 & 2 \\ 
6 & 8 & 7 & 4 & 2 & 5 & 1 & 3 & 0 \\ 
7 & 0 & 1 & 8 & 5 & 3 & 6 & 2 & 4 \\ 
8 & 5 & 0 & 2 & 1 & 6 & 4 & 7 & 3 \\ 
1 & 4 & 6 & 0 & 3 & 2 & 7 & 5 & 8 \\ 
2 & 1 & 5 & 7 & 0 & 4 & 3 & 8 & 6 \\ 
3 & 7 & 2 & 6 & 8 & 0 & 5 & 4 & 1 \\ 
4 & 2 & 8 & 3 & 7 & 1 & 0 & 6 & 5 \\ 
5 & 6 & 3 & 1 & 4 & 8 & 2 & 0 & 7%
\end{array}
}} \right)  \label{DLS 2qt}
\end{equation}

The maps given by (\ref{columnpermutation}) and (\ref{rowpermutation})
produce the following permutations: 
\begin{equation}
0\rightarrow 0\, \qquad 1\rightarrow 3\rightarrow 5\rightarrow 7\rightarrow
1\, \qquad 2\rightarrow 4\rightarrow 6\rightarrow 8\rightarrow 2.
\end{equation}
Applying these permutations to the rows and the columns of (\ref{DLS 2qt})
will return %the square of Eq.(\ref{DLS 2qt}) to standard form.
this square to standard form. 

Next, let us find the LS obtained after application of $\mathbf{X}^{2}_{1,2}$
to the commuting set defined by $f\left( \alpha \right) =\sigma ^{3}\alpha $. Given 
\begin{equation}
\mathbf{X}^{2}_{1,2}=\left( 
\begin{array}{cc}
1 & 2 \\ 
0 & 1%
\end{array}%
\right) , \qquad \mathbf{X}^{-2}_{1,2}=\left( 
\begin{array}{cc}
1 & 1 \\ 
0 & 1%
\end{array}%
\right)
\end{equation}
we obtain 
\begin{equation}
\mathbf{\Gamma}^{(g)} = (\mathbf{X}^{2}_{1,2})^{T}\mathbf{\Gamma }^{(f)}%
\mathbf{X}^{2}_{1,2} =\left( 
\begin{array}{cc}
2 & 2 \\ 
2 & 1%
\end{array}%
\right) ,
\end{equation}
which corresponds to the non-standard LS 
\begin{equation}
\tilde L^{(g)}= 
\begin{array}{c}
\\ 
\\ 
\\ 
\\ 
\\ 
\\ 
\\ 
\end{array}
\left( {\renewcommand{\arraystretch}{0.85} \renewcommand{%
\arraycolsep}{2.95pt} {\small 
\begin{array}{ccccccccc}
0 & 6 & 3 & 5 & 4 & 2 & 7 & 1 & 8 \\ 
7 & 5 & 0 & 8 & 1 & 4 & 3 & 2 & 6 \\ 
4 & 7 & 2 & 3 & 8 & 5 & 1 & 6 & 0 \\ 
6 & 2 & 8 & 4 & 7 & 0 & 5 & 3 & 1 \\ 
5 & 4 & 6 & 1 & 3 & 7 & 8 & 0 & 2 \\ 
3 & 8 & 7 & 6 & 2 & 1 & 0 & 4 & 5 \\ 
8 & 1 & 5 & 2 & 0 & 3 & 6 & 7 & 4 \\ 
2 & 0 & 1 & 7 & 5 & 6 & 4 & 8 & 3 \\ 
1 & 3 & 4 & 0 & 6 & 8 & 2 & 5 & 7%
\end{array}
}} \right)  \label{XOLS 2qt}
\end{equation}

The series of permutations that bring (\ref{XOLS 2qt}) back to the standard
form of the untransformed square, $L^{(\sigma^3 \alpha)}$, can be once again
found using Eq.~(\ref{colperm_CNOT_to_original}). There is no row
transformation. The column transformation yields 
\begin{equation}
1 \rightarrow 6 \rightarrow 4 \rightarrow 5 \rightarrow 2 \rightarrow 8
\rightarrow 1 \\
3 \to 7 \to 3
\end{equation}
% \beq
% \begin{array}{lll}
%  (\mathbf{s}^{1})^{T} &\rightarrow (\mathbf{s}^{6})^{T}&\rightarrow (\mathbf{s}^{4})^{T}\rightarrow (\mathbf{s}^{5})^{T} \rightarrow 
%  (\mathbf{s}^{2})^{T}\rightarrow (\mathbf{s}^{8})^{T}\rightarrow (%
% \mathbf{s}^{1})^{T} \\
% (\mathbf{s}^{3})^{T} &\to (\mathbf{s}^{7})^{T}&\to (\mathbf{s}^{3})^{T} 
% \end{array}
% \eeq
We must also determine the symbol transformations. Unlike in the 3 qubit
example, $c_1 = 2$ and $c_2 = 1$ so we must take into account the matrix $%
\mathbf{C}$ in our expansion, $\sigma^i = s^i_1 c_1^{-1} \theta_1 + s_2^i
c_2^{-1} \theta_2$.

Finally, using Eq.~(\ref{colperm_CNOT_to_original}) produces the equation 
\begin{equation}
\sigma^i\to (s_1^i, s_2^i) \left( 
\begin{array}{cc}
2 & 0 \\ 
0 & 1%
\end{array}
\right) \left( 
\begin{array}{cc}
1 & 2 \\ 
0 & 1%
\end{array}
\right) \left( 
\begin{array}{cc}
2 & 0 \\ 
0 & 1%
\end{array}
\right) \left( 
\begin{array}{c}
\theta_1 \\ 
\theta_2%
\end{array}
\right)
\end{equation}
which leads to the symbol permutations 
\begin{equation}
\sigma^1 \to \sigma^3 \to \sigma^8 \to \sigma^1 \\
\sigma^4 \to \sigma^5 \to \sigma^7 \to \sigma^4 \\
\end{equation}
meaning we must perform the symbol swaps $1 \to 3 \to 8 \to 1$ and $4 \to 5
\to 7 \to 4$.

For a single qutrit, there are eight generalized Paulis: $\mathcal{Z}, 
\mathcal{X}, \mathcal{Z} \mathcal{X}$, $\mathcal{Z}^2 \mathcal{X}$, and
their squares. Under the local transformation 
\begin{equation}
\mathbf{U}^S_1 = \left( {\renewcommand{\arraystretch}{0.85} %
\renewcommand{\arraycolsep}{2.95pt} {\small 
\begin{array}{ccc}
1&  0 & 0 \\ 
0  & 0 & 1\\ 
0 & 1 &  0
\end{array}
}} \right)
\end{equation}
they are mapped (up to a phase) to 
\begin{equation}
\mathcal{Z} \leftrightarrow \mathcal{Z}^2\, , \quad \mathcal{X}
\leftrightarrow \mathcal{X}^2 \, ,\quad \mathcal{ZX} \leftrightarrow (%
\mathcal{ZX})^2\, ,\quad (\mathcal{Z}^2\mathcal{X}) \leftrightarrow (%
\mathcal{Z}^2\mathcal{X})^2\, . \label{Paulimapping}
\end{equation}

Choose the generating set $G = \{2\sigma^4,\sigma^2\}=\{ \sigma^8,\sigma^2 \}$, and the curve 
$f(\alpha) = \sigma^4 \alpha$. Then: 
\begin{equation}
\mathbf{A}^{(\sigma^4 \alpha)} = \left( 
\begin{array}{cc|cc}
1 & 0 & 1 & 0 \\ 
0 & 1 & 0 & 2%
\end{array}
\right)\, .
\end{equation}

The monomials corresponding to each generator are: 
\begin{table}[h!]
\centering
\begin{tabular}{c|c|c}
Generator & Monomial & Pauli representation \\ \hline
$g_1 = \sigma^8$ & $Z_{\sigma^8} X_{\sigma^4}$ & $\mathcal{ZX} \otimes %
\mathinner{\hbox{1}\mkern-4mu\hbox{l}} $ \\ 
$g_2 = \sigma^2$ & $Z_{\sigma^2} X_{\sigma^6}$ & $\mathinner{\hbox{1}%
\mkern-4mu\hbox{l}} \otimes (\mathcal{Z}\mathcal{X}^2)$%
\end{tabular}%
\end{table}

The transformation $\mathbf{\ U}^S_1$ corresponds to the $2 \times 2$ map 
\begin{equation}
T(\mathbf{U}^S_1) = \left( {\renewcommand{\arraystretch}{0.85} %
\renewcommand{\arraycolsep}{2.95pt} {\small 
\begin{array}{cc}
2 & 0 \\ 
0 & 2%
\end{array}
}} \right)  \label{tu}
\end{equation}
shuffling the powers $a_1,b_1$ in the initial monomial. Now, $\mathbf{U} = 
\mathbf{U}^S_1 \otimes \mathinner{\hbox{1}\mkern-4mu\hbox{l}}$. Given $T(%
\mathinner{\hbox{1}\mkern-4mu\hbox{l}}) = \mathinner{\hbox{1}\mkern-4mu%
\hbox{l}}$, we readily obtain $\mathbf{K}_s$: 
\begin{equation}
{\renewcommand{\arraystretch}{2.0} 
\begin{array}{ccc}
\mathbf{K}_{11} = \left( {\renewcommand{\arraystretch}{0.85} %
\renewcommand{\arraycolsep}{2.95pt} {\small 
\begin{array}{cc}
2 & 0 \\ 
0 & 1%
\end{array}
}} \right)\, ,\  & \quad & \mathbf{K}_{12} = \left( {\renewcommand{%
\arraystretch}{0.85} \renewcommand{\arraycolsep}{2.95pt} {\small 
\begin{array}{cc}
0 & 0 \\ 
0 & 0%
\end{array}
}} \right) \, , \\ 
\mathbf{K}_{21} = \left( {\renewcommand{\arraystretch}{0.85} %
\renewcommand{\arraycolsep}{2.95pt} {\small 
\begin{array}{cc}
0 & 0 \\ 
0 & 0%
\end{array}
}} \right)\, , &  & \mathbf{K}_{22} = \left( {\renewcommand{%
\arraystretch}{0.85} \renewcommand{\arraycolsep}{2.95pt} {\small 
\begin{array}{cc}
2 & 0 \\ 
0 & 1%
\end{array}
}} \right) \, ,%
\end{array}
}
\end{equation}
so that 
\begin{equation}
\mathbf{\tilde{A}}^{(f^\prime)}= \left( 
\begin{array}{cc|cc}
1 & 0 & 1 & 0 \\ 
0 & 1 & 0 & 2%
\end{array}
\right) \left( {\renewcommand{\arraystretch}{0.85} \renewcommand{%
\arraycolsep}{2.95pt} {\small 
\begin{array}{cccc}
2 & 0 & 0 & 0 \\ 
0 & 1 & 0 & 0 \\ 
0 & 0 & 2 & 0 \\ 
0 & 0 & 0 & 1%
\end{array}
}} \right) = \left( 
\begin{array}{cc|cc}
2 & 0 & 2 & 0 \\ 
0 & 1 & 0 & 2%
\end{array}
\right)
\end{equation}

It follows from Eq. (\ref{Paulimapping}) that the new generating elements are
\begin{table}[h!]
\centering
\begin{tabular}{c|c|c}
Pauli representation & Monomial & Generator \\ \hline
$\left(\mathcal{ZX}\right)^2 \otimes \mathinner{\hbox{1}\mkern-4mu\hbox{l}} $
& $Z_{\sigma^4} X_{\sigma^8}$ & $g_1^\prime = \sigma^4$ \\ 
$\mathinner{\hbox{1}\mkern-4mu\hbox{l}} \otimes \left(\mathcal{Z}\mathcal{X}%
^2\right) $ & $Z_{\sigma^2} X_{\sigma^6}$ & $g_2^\prime = \sigma^2$%
\end{tabular}%
\end{table}

Using Eqs. (\ref{Gf_as_GaGb}) and (\ref{A_as_ga_gb}), we can transform this
to the standard form $(\,\mathinner{\hbox{1}\mkern-4mu\hbox{l}}\, |\, 
\mathbf{\Gamma}^{(f^\prime)}\, )$: 
\begin{equation}
\mathbf{A}^{(f^\prime)} = \left( 
\begin{array}{cc|cc}
1 & 0 & 1 & 0 \\ 
0 & 1 & 0 & 2%
\end{array}
\right),
\end{equation}
and we can see that the resultant curve is both invertible and commutative,
as the adjacency matrix is symmetric and det$[\mathbf{\Gamma}^{(f^\prime)}]
\neq 0$. In this case, it just so happens that the curve is not changed
under transformation; this is not always the case.

\end{document}